%% file: main.tex
\documentclass[a4paper]{article}

\pdfoutput=1

\usepackage[english]{babel}

\usepackage[protrusion=true,expansion=true]{microtype}
\usepackage[T1]{fontenc}

\usepackage{lmodern}

\usepackage{mathtools}
\usepackage{amssymb}

\usepackage{bm} 
\usepackage{upgreek} 
\usepackage{bbold} 

\usepackage{textcomp} 
\usepackage{gensymb} 

\usepackage{graphicx} 
\usepackage{float} 
\usepackage{placeins}
\usepackage{hyperref} 
\usepackage[margin=2.5cm]{geometry} 
\usepackage[labelfont=bf]{caption} 
\usepackage{subcaption}
\usepackage{authblk}                 

\usepackage{color} 
\usepackage{physics} 
\usepackage{rotating} 
\usepackage{longtable} 
\usepackage{cleveref} 

\usepackage{csquotes} 

\usepackage{makecell}
\usepackage{booktabs}
\usepackage{multirow}

\usepackage[style=numeric-comp,url=true,sorting=none]{biblatex}
\renewbibmacro*{doi+eprint+url}{%
    \printfield{doi}%
    \newunit\newblock%
    \iftoggle{bbx:eprint}{%
        \usebibmacro{eprint}%
    }{}%
    \newunit\newblock%
    \iffieldundef{doi}{%
        \usebibmacro{url+urldate}}%
        {}%
    }
\newcommand{\N}{\ensuremath{\mathbb{N}}}

\newcommand{\R}{\ensuremath{\mathbb{R}}}
\newcommand{\C}{\ensuremath{\mathbb{C}}}
\newcommand{\bmat}[1]{\ensuremath{\begin{bmatrix}#1\end{bmatrix}}}

\newcommand{\calH}{\ensuremath{\mathcal{H}}}
\newcommand{\calS}{\ensuremath{\mathcal{S}}}
\newcommand{\calC}{\ensuremath{\mathcal{C}}}
\newcommand{\kv}[1]{\ensuremath{\ket{\vec{#1}}}}

\DeclareMathOperator*{\argmin}{arg\,min}

\title{Quantum Computing and Tensor Networks for Laminate Design: A Novel Approach to Stacking Sequence Retrieval}
\author[1]{Arne Wulff}
\author[1]{Boyang Chen\footnote{Corresponding author, \href{mailto:b.chen-2@tudelft.nl}{b.chen-2@tudelft.nl}}}
\author[2]{Matthew Steinberg}
\author[1]{Yinglu Tang}
\author[2]{Matthias M\"oller}
\author[2]{Sebastian Feld}
\affil[1]{Faculty of Aerospace Engineering, Delft University of Technology, The Netherlands}
\affil[2]{Faculty of Electrical Engineering, Mathematics and Computer Science, Delft University of Technology, The Netherlands}
\date{}

\begin{document}
\maketitle

{\abstract{\noindent\input{abstract}}}


\input{I_introduction.tex}

\input{II_stackingretrieval.tex}

\input{III_quantum}
\input{IV_algos}
\input{V_demo}

\input{VI_conclusion}

\input{formalities}

\cleardoublepage
\input{appendix}

\cleardoublepage

\printbibliography[title=References]

\end{document}

%% file: abstract.tex
As with many tasks in engineering, structural design frequently involves navigating complex and computationally expensive problems. A prime example is the weight optimization of laminated composite materials, which to this day remains a formidable task, due to an exponentially large configuration space and non-linear constraints. The rapidly developing field of quantum computation may offer novel approaches for addressing these intricate problems. However, before applying any quantum algorithm to a given problem, it must be translated into a form that is compatible with the underlying operations on a quantum computer.
Our work specifically targets stacking sequence retrieval with lamination parameters, which is typically the second phase in a common bi-level optimization procedure for minimizing the weight of composite structures. To adapt stacking sequence retrieval for quantum computational methods, we map the possible stacking sequences onto a quantum state space. We further derive a linear operator, the Hamiltonian, within this state space that encapsulates the loss function inherent to the stacking sequence retrieval problem. Additionally, we demonstrate the incorporation of manufacturing constraints on stacking sequences as penalty terms in the Hamiltonian. This quantum representation is suitable for a variety of classical and quantum algorithms for finding the ground state of a quantum Hamiltonian. For a practical demonstration, we performed numerical state-vector simulations of two variational quantum algorithms and additionally chose a classical tensor network algorithm, the DMRG algorithm, to numerically validate our approach. For the DMRG algorithm, we derived a matrix product operator representation of the loss function Hamiltonian and the penalty terms. Although this work primarily concentrates on quantum computation, the application of tensor network algorithms presents a novel quantum-inspired approach for stacking sequence retrieval.

%% file: I_introduction.tex
\section{Introduction}

Across various engineering disciplines, researchers and practitioners regularly encounter complex and computationally intensive problems \cite{Zhang2010,Roy2008,Karp1972}. The rapidly evolving field of quantum computing could offer crucial breakthroughs in addressing these issues and exceed the capabilities of conventional methods \cite{Gill2022,Lu2023,Kim2023,Nielsen2012,Feynman1982,qcprogress2019}. As this technology continues to mature, the potential benefits it offers to complex engineering problems are becoming more apparent \cite{Ajagekar2019,McArdle2020,Paudel2022,Giani2021,Bauer2020,Andersson2022,Liu2022,Wang2023,Alexeev2023,Osaba2022,Mohseni2022,Egger2020,Herman2022,Egger2021}. However, the challenge lies in accurately identifying those scenarios where quantum computing holds an advantage and in crafting algorithms to harness this potential. To address this, initiating the development of quantum algorithms for engineering applications now -- in parallel with quantum hardware advancements -- is essential. Such an approach positions us to immediately harness the full capabilities of quantum computing in engineering as soon as it becomes available. 

The design of laminated composite materials \cite{Ghiasi2009,Ghiasi2010} is an example of a complex engineering challenge which could benefit from a quantum-based approach. These materials are extensively used in aerospace applications for their high strength-to-weight ratio. Furthermore, these materials consist of many configurable layers, each impacting the overall stiffness. The stiffness characteristics of the whole material can therefore be tailored to the specific requirements of a structure \cite{Nikbakt2018,Nikbakht2019}. However, determining the optimal arrangement for specific applications from the exponentially large configuration space is a significant challenge, positioning laminated composites as a suitable candidate for quantum computing applications \cite{IBMBoeing2023,Fuller2021}.

A widely used method in designing composite structures involves a bi-level optimization procedure \cite{Yamazaki1996,Herencia2007,IJsselmuiden2009,Liu2011,Liu2013,Macquart2017,Liu2019}. The initial phase focuses on optimizing the material's thickness and stiffness characteristics, expressed through lamination parameters \cite{Tsai1968,Miki1991}. Subsequently, the second phase involves finding a feasible stacking sequence that aligns with the lamination parameters determined earlier. While the first phase typically employs continuous gradient descent methods, the second stage, stacking sequence retrieval, constitutes an intricate combinatorial optimization problem, particularly when incorporating manufacturing considerations \cite{Niu1988,Bailie1997,compmathandbook2002,Albazzan2019}. 

Optimizing stacking sequences for constant stiffness structures has been addressed through various methods, including layer-wise optimization \cite{Narita2003,vanCampen2009}, branch-and-bound \cite{Todoroki2004,Terada2001,Todoroki2007,Liu2019} and beam search \cite{Fedon2021} techniques. The complexity increases substantially for structures that necessitate variable stiffness, often achieved by dividing the structure into distinct panels \cite{Kim1999,Soremekun2002}, each requiring a specific stacking sequence and stiffness. Ensuring structural integrity by matching the stacking sequences across different panels, in accordance with blending rules \cite{Kristinsdottir2001,Adams2004,Zabinsky2006,vanCampen2008}, adds a further layer of complexity, and adapting the previously mentioned methods to this multi-panel scenario is challenging. 
Given this complexity, often intensified by the curse of dimensionality, current methods generally rely on meta-heuristics such as genetic algorithms \cite{Todoroki1998,Autio2000,Herencia2007,Bloomfield2008,IJsselmuiden2009,Liu2011,Macquart2017,Liu2019} or particle swarm optimization \cite{Bloomfield2008,Bloomfield2009,Bloomfield2010}, which are frequently used in conjunction with problem-specific methods like stacking sequence tables \cite{Irisarri2014,Meddaikar2017} to address these challenges. Despite this progress, the quest for more efficient solutions in stacking sequence retrieval remains a pivotal area of research, indicating the necessity for novel approaches in this complex domain.
Quantum computing may offer a unique approach to manage these complex, high dimensional scenarios effectively. The formulation of these algorithms however and their implementation on contemporary quantum hardware remain a challenging tasks. Before tackling complex multi-panel cases, it is essential to gain a deeper understanding of simpler scenarios, where the limitations of existing technology are less constraining. Therefore, and despite the existence of effective stacking sequence retrieval algorithms for single constant-stiffness panels, the foundational work for quantum algorithms must also begin with this simpler case.

In this work, we explore the adaptation of the \textit{stacking sequence retrieval problem} (SSR) with lamination parameters for constant stiffness panels to quantum algorithms. Initially, we formalize the problem as an integer-optimization problem, suitable for translation into a quantum mechanics framework. This involves representing variable configurations of the problem as states in a Hilbert space and deriving a Hamiltonian to embody the problem's loss function. The objective then becomes identifying the ground-state of the Hamiltonian, a task addressable by various quantum algorithms \cite{Cao2019,McArdle2020,Ayral2023,Peruzzo2014,Farhi2014,Cerezo2021,Endo2021,Kadowaki1998,Kadowaki2002,Hauke2020,Kitaev1995,Abrams1999,Kim2023}. We also present how manufacturing constraints can be incorporated as penalty terms within the Hamiltonian. As a demonstration of our formalism, we selected two variational quantum algorithms, the quantum approximate optimization algorithm (QAOA) \cite{Farhi2014,Zhou2020} and a hardware efficient approach, for which we perform state-vector simulations. Additionally, we chose the density matrix renormalization group algorithm (DMRG), a classical tensor network method effective in finding ground states of many-body systems \cite{White1992,White1993,Nishino1996,Verstraete2004,Schollwock2011,Chan2016}. For this purpose, the Hamiltonian must first be expressed in terms of a matrix product operator (MPO), which is a particular tensor network structure. This enables us to optimize systems as large as 200 plies, by controlling the bond dimension of the states' MPS.
For all three algorithms, we conducted tests under various conditions dependent on the algorithm at hand. The optimization results successfully approximated the optimal solutions, validating our Hamiltonian construction and confirming the efficacy of our approach. Furthermore, the notable performance of the DMRG algorithm points to the potential of tensor network methods as a quantum-inspired alternative to conventional stacking sequence retrieval methods.

Following this introduction, \cref{sec:stackingretrieval} begins with an overview of laminated composite materials and concludes with the formulation of the specific integer-optimization problem that we employ in the remainder of this work. In \cref{sec:quantum}, we map the problem to a quantum state space, and construct the Hamiltonian corresponding to the problem's loss function. We additionally show how to integrate manufacturing constraints as penalty terms in the Hamiltonian. In \cref{sec:algo}, we discuss the algorithms that we selected for  numerical demonstrations. In particular, this section discusses how the Hamiltonian must be adapted for QAOA and the DMRG algorithm. The numerical demonstration can be found in \cref{sec:demo}. The paper concludes with \cref{sec:conclusion}, and summarizes our findings, as well as ramifications for the field of quantum computing in the context of engineering and materials science applications. 

%% file: II_stackingretrieval.tex
\section{Stacking sequence retrieval as an optimization problem}\label{sec:stackingretrieval}

\subsection{Laminated composite materials}

\begin{figure}
    \centering
    \includegraphics[scale=1.0]{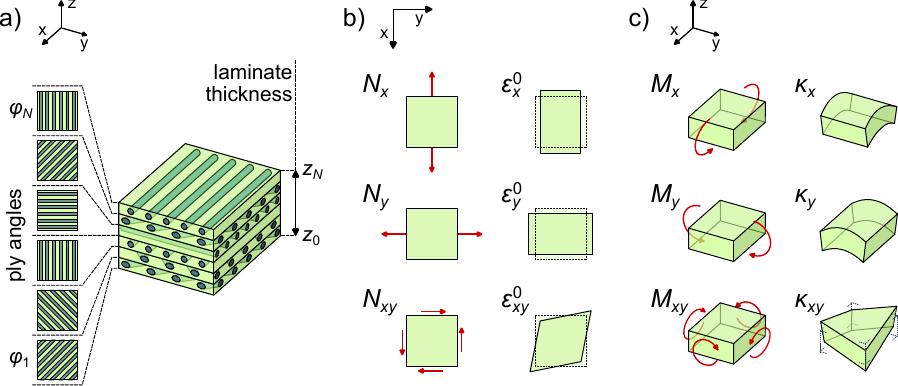}
    \caption{a) Diagram of a laminated composite material, which consists of multiple layers that are oriented in different directions. b) A representation of the stress resultants $\vec{N}$, which for isotropic materials result in in-plane deformations $\vec{\varepsilon}^{\: 0}$. c) A pictographical representation of the moment resultants $\vec{M}$, which for isotropic materials result in out-of-plane bending $\vec{\kappa}$. Non-diagonal elements in the $\mathbf{ABD}$-matrix allow for coupling between all possible stress and moment resultants on the one hand, and all in- and out-of-plane deformations on the other hand.}
    \label{fig:laminate}
\end{figure}

\textit{Fiber-reinforced composite materials} are heterogeneous materials consisting of fibers incorporated in a supportive matrix, which gives rise to an orthotropic material with distinct stiffness characteristics along the longitudinal and the orthogonal fiber-directions \cite{fibertechnology2017}. \textit{Laminated composites} are formed by stacking multiple layers of this material, for which the anisotropic stiffness characteristics can be tailored to the unique strength requirements of a structure by adjusting the orientation of each ply (\Cref{fig:laminate} a).

For laminated composites, the mechanical behavior, to a first-order approximation, is typically characterized using a variation of Hooke's law \cite{Reissner1961,Dong1962,Tsai1980}:
\begin{equation}
    \begin{bmatrix}
    \vec{N} \\ \vec{M}
\end{bmatrix} = \begin{bmatrix}
    \mathbf{A} & \mathbf{B} \\ \mathbf{B} & \mathbf{D}
\end{bmatrix} \begin{bmatrix}
    \vec{\varepsilon}^{\: 0} \\ \vec{\kappa}
\end{bmatrix}.
\end{equation}
Here, the stress and moment resultants, $\vec{N}$ and $\vec{M}$, represent the loadings within the structure. The deformations are characterized by $\vec{\varepsilon}^{\: 0}$ (in-plane bending) and $\vec{\kappa}$ (out-of-plane bending). Figures \ref{fig:laminate} b and c illustrate the stress and moment resultants, along with the deformations, visually demonstrating their effects on the laminate. The $\mathbf{ABD}$-matrix analogously functions as the stiffness tensor. Specifically, $\mathbf{A}$ and $\mathbf{D}$ define the in-plane and out-of-plane stiffness, respectively. A non-zero $\mathbf{B}$ indicates a coupling between them, meaning in-plane stresses can result from out-of-plane bending and vice versa. Both vectors $\vec{N}$ and $\vec{M}$ along with $\vec{\varepsilon}^{\: 0}$ and $\vec{\kappa}$ consist of three components. Furthermore, matrices $\mathbf{A}$, $\mathbf{B}$ and $\mathbf{D}$ are of dimensions $3\times3$.

For a laminate defined by a specific stacking sequence, the $\mathbf{ABD}$-matrix is completely determined by the thickness of the material, the angles of the individual plies, and the inherent properties of the selected composite material. Furthermore, these dependencies separate when introducing \textit{lamination parameters}, which distill the complete ply-angle dependence of the $\mathbf{ABD}$-matrix into exactly 12 real values $v_l^A,v_l^B,v_l^D$ with $l = 1,2,3,4$ \cite{Tsai1968,Miki1991}. Denoting the laminate thickness with $h$, the $\mathbf{ABD}$-matrix is determined by \cite{IJsselmuiden2011,Albazzan2019}:
\begin{align} \label{eq:abd_from_lp}
\mathbf{A} &= h \qty(\bm{\Gamma}_0 + \bm{\Gamma}_1 v_1^A + \bm{\Gamma}_2 v_2^A + \bm{\Gamma}_3 v_3^A + \bm{\Gamma}_4 v_4^A),\\
\mathbf{B} &= \frac{h^2}{4} \qty(\bm{\Gamma}_0 + \bm{\Gamma}_1 v_1^B + \bm{\Gamma}_2 v_2^B + \bm{\Gamma}_3 v_3^B + \bm{\Gamma}_4 v_4^B), \notag\\
\mathbf{D} &= \frac{h^3}{12} \qty(\bm{\Gamma}_0 + \bm{\Gamma}_1 v_1^D + \bm{\Gamma}_2 v_2^D + \bm{\Gamma}_3 v_3^D + \bm{\Gamma}_4 v_4^D), \notag
\end{align}
where the intrinsic material properties enter the equation in form of the matrices $\Gamma_l$ with $l=0,1,2,3,4$. For a stacking sequence comprising $N$ plies of equal thickness and ply angles $\phi_1,\dots,\phi_N \in (-\pi/2,\pi/2]$, and the distance of ply $n$ from the mid-plane captured in \(z_n = -N/2 + n\), the lamination parameters read:
\begin{align} \label{eq:lp}
    (v_1^A,v_2^A,v_3^A,v_4^A) &= \frac{1}{N} \sum_{n=1}^N \qty(z_n - z_{n-1}) \qty(\cos(2\phi_n),\sin(2\phi_n),\cos(4\phi_n),\sin(4\phi_n)),\\
     (v_1^B,v_2^B,v_3^B,v_4^B) &= \frac{2}{N^2} \sum_{n=1}^N \qty(z_n^2 - z_{n-1}^2) \qty(\cos(2\phi_n),\sin(2\phi_n),\cos(4\phi_n),\sin(4\phi_n)), \notag \\
      (v_1^D,v_2^D,v_3^D,v_4^D) &= \frac{4}{N^3} \sum_{n=1}^N \qty(z_n^3 - z_{n-1}^3) \qty(\cos(2\phi_n),\sin(2\phi_n),\cos(4\phi_n),\sin(4\phi_n)). \notag
\end{align}
The matrices $\bm{\Gamma}_k$ are constructed with the Tsai-Pagano material invariants \(U_1, \dots, U_5 \in \R\) \cite{Tsai1968}:
\begin{equation}
    \bm{\Gamma}_0 = \begin{bmatrix}
    U_1 & U_4 & 0  \\
    U_4 & U_1 & 0 \\
    0 & 0 & U_5
\end{bmatrix}, \quad \bm{\Gamma}_1 = \begin{bmatrix}
    U_2 & 0 & 0  \\
    0 & -U_2 & 0 \\
    0 & 0 & 0
\end{bmatrix}, \quad \bm{\Gamma}_2 = \begin{bmatrix}
    0 & 0 & U_2/2 \\
    0 & 0 & U_2/2 \\
    U_2/2 & U_2/2 & 0 
\end{bmatrix}
\end{equation}
\begin{equation*}
    \bm{\Gamma}_3 = \begin{bmatrix}
    U_3 & -U_3 & 0  \\
    -U_3 & U_3 & 0 \\
    0 & 0 & -U_3
\end{bmatrix}, \quad \bm{\Gamma}_4 = \begin{bmatrix}
     0 & 0 & U_3 \\
     0 & 0 & -U_3 \\
    U_3 & -U_3 & 0
\end{bmatrix}.
\end{equation*}
These material invariants depend only on the longitudinal, transverse, and shear moduli, as well as the Poisson's ratio of the composite.

In numerous applications, particularly in aerospace structural design, materials are required to be as lightweight as possible while simultaneously possessing the strength to withstand forces and stresses under a variety of expected loading conditions. A typical objective involves minimizing the thickness or, equivalently, the number of plies in composite materials. This minimization is constrained by the expected stress and moment resultants $\vec{N}$ and $\vec{M}$, alongside predefined failure criteria expressed through inequalities involving stresses, deformations, or the $\mathbf{ABD}$-matrix. The optimization process becomes highly complex due to the non-linear relationship between the laminate's structural behavior and the ply angles, which serve as design variables in the optimization and typically range from tens to hundreds in number \cite{IJsselmuiden2011}. To effectively navigate this complexity, a bi-level design strategy is commonly employed for laminated composite materials, involving two distinct phases of optimization \cite{Yamazaki1996,Herencia2007,IJsselmuiden2009,Liu2011,Liu2013,Macquart2017,Liu2019}.
The first phase optimizes for the laminate thickness $h$ and the stiffness characteristics, represented by lamination parameters $v_l^A$, $v_l^B$ and $v_l^D$, such that they meet predefined strength criteria. These design variables are typically assumed to be continuous, such that gradient descent methods can be employed effectively. The subsequent phase, termed \textit{stacking sequence retrieval} (SSR), seeks a ply-angle sequence that approximates the lamination parameters derived from the first phase. The laminate thickness $h$ is incorporated into the second phase as the total number of plies $N$ in the laminate. In this work, we will focus on this second phase of stacking sequence retrieval.

Manufacturing constraints typically dictate the allowed ply-angle sequences in a laminate. Primary among these is the confinement of ply angles to a discrete set, with a common selection being $\{0^\circ, +45^\circ, 90^\circ, -45^\circ\}$. Alternatives can also be found, with divisions of the interval $(-90^\circ, 90^\circ]$ at intervals of $22.5^\circ$ or $15^\circ$. It is worth noting that fiber orientations, and consequently the lamination parameters, demonstrate a periodicity of $180^\circ$. Other prevalent manufacturing constraints include \cite{Niu1988,Bailie1997,compmathandbook2002,Albazzan2019}:
\begin{itemize}
\item \textbf{Symmetric laminates:} The laminate exhibits symmetry about its midplane. Hence, plies $n$ and $N - n + 1$ share the same ply angle, leading to null coupling lamination parameters $v_k^B$ and stiffness matrix $\mathbf{B}$.
\item \textbf{Disorientation constraint:} The angular difference between adjacent plies is limited, often to $45^\circ$.
\item \textbf{Balanced laminates:} Except for ply-angles $0^\circ$ or $90^\circ$, the number of plies with angles $\theta$ and $-\theta$ are equal. This has the effect, that $v_2^A$ and $v_4^A$ vanish.
\item \textbf{Contiguity constraint:} The number of consecutive same-angle plies is limited, for instance to 5 plies with the same orientation.
\item \textbf{10\%-rule:} Specific ply angles must constitute a minimum of 10\% of the laminate. This rule is frequently applied to angles of $0\degree$, $90\degree$ and $\pm 45\degree$.
\end{itemize}

For a laminate consisting of $N$ plies and a set of $d$ discrete angles, the solution space encompasses $d^N$ potential configurations. This exponential explosion renders direct search or enumeration methods computationally prohibitive. Moreover, the imposition of manufacturing constraints further complicates the search landscape. Various strategies have been advanced to navigate this complex space \cite{Ghiasi2009,Nikbakt2018,Fedon2021,Sprengholz2021}, however, the optimization of more complicated structures remains a formidable challenge. 

Lately, there has been a spotlight on breakthroughs in quantum computing, a technology that might offer new tools to address this complexity \cite{Mosca2008,Montanaro2016,Nielsen2012,Bharti2022,Lu2023}.
However, one must map the presently-described problem to a representation amenable to quantum information-theoretic methods.
We start with formulating stacking sequence retrieval as an integer-optimization problem.

\subsection{Formulating stacking sequence retrieval as an integer-optimization problem} \label{sec:ssrinteger}

In general, it is advantageous for optimization algorithms to be presented with a version of the problem that abstracts away all but the essential information. For our purposes, this level of abstraction is necessary to adapt the problem for quantum computing architectures. In this section, we present our redefinition of the problem that forms the basis for our analysis in this work.

The discrete nature of the ply angles in laminated composites is well-suited to be represented in terms of integer design variables, a method commonly employed in various stacking sequence retrieval methods \cite{Autio2000,Liu2019,Ntourmas2021}. Accordingly, we define the $d$ discrete angles $\Theta = \{\theta_1, \theta_2, \dots, \theta_d\}$, allowing each angle $\theta_s$ to be represented by its corresponding label $s \in \{1, \dots, d\}$, effectively using these labels as integer design variables.
Consequently, we can identify the ply angle sequence $\vec{\theta} = (\theta_{s_1},\theta_{s_2},\dots,\theta_{s_N})$ with an ordered list of integers $\vec{s}=(s_1,s_2,\dots,s_N)$. Here we use $\theta$ instead of the $\phi$ from \cref{eq:lp}, to convey that we now explicitly choose the ply angles from the discrete set $\Theta$. In the following, we denote the set of all possible stacking sequences as $\mathcal{S} = \{1,\dots,d\}^N$, such that $\vec{s} \in \mathcal{S}$.  Considering \cref{eq:lp}, the lamination parameters can then be written in a common general form:
\begin{equation} \label{eq:lp_gen}
    v^X_l(s_1,\dots,s_N) = \sum_{n=1}^N \alpha_n^X f_l(s_n), 
\end{equation}
where $X = A,B,D$ and  $l = 1,2,3,4$, and the functions $f_l:\{1,\dots,d\} \to [-1,1]$ defined as:
\begin{align}
    f_1(s) &= \cos(2 \theta_s), & f_2(s) &= \sin(2 \theta_s), \\
    f_3(s) &= \cos(4 \theta_s), & f_4(s) &= \sin(4 \theta_s). \notag
\end{align}
In this framework, a manufacturing constraint can be realized as a Boolean function $c:\mathcal{S}\to\{\mathrm{True},\mathrm{False}\}$ on the lists of integers, where $c(\vec{s})=\mathrm{True}$ exactly when ply-angle sequence $\vec{\theta} = (\theta_{s_1},\dots,\theta_{s_N})$ satisfies the constraint. The optimization towards target lamination parameters can then be written as an integer-optimization problem. We choose an abstracted formulation of the problem that does not include the ply angles $\theta_s$ but assumes their impact on the loss function to be encoded in the functions $f_l$ that are defined on the according indices $s = \{1,\dots,d\}$:
\begin{quote}\textbf{Stacking sequence retrieval problem (SSR):} Given
\begin{itemize}
    \item Positive integers $N,d \in \N$,
    \item The set of stacking sequences $\mathcal{S}=\{1,\dots,d\}^N$,
    \item Sets of weights $\{\alpha_n^X\}_n$ for $n=1,\dots,N$ and $X=1,\dots,N_\alpha$ for some $N_\alpha \in \N$,
    \item Functions $f_l: \{1,\dots,d\} \to \R$ for $l = 1,...,N_f$ for some $N_f \in \N$,
    \item A set of constraints $\mathcal{C}=\qty{c_k:\mathcal{S}\to \{\mathrm{True},\mathrm{False}\}}$
    on the solution space, where $k$ enumerates the constraints,
    \item Target values $\vec{\xi} \in \R^{N_\alpha N_f}$,
\end{itemize}
find a stacking sequence $\vec{s} = (s_1,\dots,s_N) \in \mathcal{S}$ with $c(\vec{s}) = \mathrm{True}$ for all $c \in \mathcal{C}$ that minimizes the loss function:
\begin{equation}
H(s_1,\dots,s_N) = \sum_{X,l} \qty(\sum_{n=1}^N \alpha_n^X f_l(s_n) - \xi^X_l)^2. \label{eq:loss_func}
\end{equation}
\end{quote}
Note, that for the loss function, we use the mean-square-error (MSE) rather than the root-mean-square-error (RMSE), due to its natural implementation in the quantum context. Nevertheless, the RMSE can prove useful to evaluate the quality of solutions, as it represents the Euclidean distance of the solution to the target parameters in lamination parameter space. The defined integer-optimization problem is more general than the original stacking sequence retrieval problem, and most of the following discussion does not depend on explicit definitions of $\alpha_n^X$, $N_\alpha$, $f_l$ and $N_f$. However, we will generally assume the previously defined case of $N_f=4$, $X=A,B,D$ such that $N_\alpha = 3$, and $\alpha_n^X$ and $f_l$ defined according to the lamination parameters in \cref{eq:lp}.

%% file: III_quantum.tex
\section{A formulation of the SSR problem in a quantum state space} \label{sec:quantum}

In this section, we discuss how the stacking sequence retrieval problem can be represented with linear algebra concepts inline with quantum information theoretic methods. In particular, we represent stacking sequences as vectors in a complex-valued Hilbert space, and treat functions acting on stacking sequences, such as the loss function or constraints, as Hermitian operators on this space. The presented approach is applicable to various prevalent quantum computing methods,  such as variational quantum algorithms (VQA) \cite{Peruzzo2014,Farhi2014,Cerezo2021,Endo2021} and quantum annealing \cite{Hauke2020,Kadowaki1998,Kadowaki2002,Neven2008,Bian2010,Rieffel2015}. For readers unfamiliar with quantum computation, we recommend the book \cite{Nielsen2012}. 

\subsection{SSR in terms of quantum states and operators}

In the defined integer optimization problem in \cref{sec:ssrinteger}, the three primary elements are the stacking sequences $\vec{s} \in \calS$, the loss function $H(\vec{s})$ in \cref{eq:loss_func}, and the constraints $c \in \calC$. Initially, we focus on the stacking sequences and loss function, and address the constraints later on.

We encode stacking sequences $\vec{s} \in \calS$ using a basis-encoding within a composite complex-valued Hilbert space $\calH = \bigotimes_{n=1}^N \calH_n$. Here, each subsystem $\calH_n$ with dimension $\dim(\calH_n) = d$ represents a single ply, and consequently the composite system for the entire stacking sequence has dimension $\dim(\calH) = d^N$. Accordingly, each basis vector $\kv{s} \in \calH$ represents an individual stacking sequence $\vec{s} \in \calS$. These vectors are orthonormal, meaning for any two distinct sequences $\vec{s}, \vec{t} \in \calS$, their inner product satisfies:
\begin{equation}
    \ip{\vec{s}}{\vec{t}} = \delta_{\vec{s},\vec{t}}.
\end{equation}

To model the loss function $H(\vec{s})$ mentioned in \cref{eq:loss_func}, we introduce a Hermitian operator $\hat{H}$ in Hilbert space $\calH$, termed the loss function Hamiltonian. This Hamiltonian is defined to be diagonal in the basis of stacking sequences with eigenvalues corresponding to the loss function evaluations $\hat{H} \kv{s} = H(\vec{s}) \kv{s}$:
\begin{equation}
    \hat{H} = \sum_{\vec{s}\in\calS} H(\vec{s}) \dyad{s}.
\end{equation}
Thus, finding the solution to the integer optimization problem is equivalent to identifying the ground state of this Hamiltonian, ignoring any additional constraints $c \in \calC$ for the moment.

Due to its generality, the basis-state encoding together with the loss function Hamiltonian is applicable to various quantum methods. However, the specific form of the Hamiltonian and its application in quantum algorithms significantly influence the practical implementation.  As we will discuss for the QAOA algorithm in \cref{sec:qaoa}, some algorithms require implementing the exponential of the Hamiltonian using the available gates on the quantum computer \cite{Farhi2014}. For feasible implementation with a limited number of quantum gates, the Hamiltonian should exhibit a low-degree polynomial expansion in terms of local operators on each subsystem.  As a further example, quantum annealing requires the polynomial expansion of the Hamiltonian to be quadratic, since the individual terms are implemented as interactions between pairs of subsystems \cite{Kadowaki1998,Kadowaki2002,Hauke2020}. The choice of mean-square-error for the loss function $H(\vec{s})$ in \cref{eq:loss_func} naturally results in a quadratic expression in the single-ply functions $f_l:\{1,\dots,d\}\to\R$, aligning with algorithmic requirements. 

To make the polynomial expansion of the loss function and the according Hamiltonian explicit, as well as for later convenience, we absorb the target parameters $\xi^X_l$ into the sum over the ply indices $n$, such that the loss function takes the form:
\begin{equation}
    H(\vec{s}) = \sum_{X,l}\qty(\sum_{n=1}^N H_{X,l}^{[n]}(s_n))^2.
\end{equation}
where the target parameters are incorporated into the ply-angle-dependent terms $H_{X,l}^{[n]}: \{1,\dots,d\} \to \R$. For the lamination parameters $X=A,B,D$, the summation over the absolute values of the weights is $\sum_{n=1}^N \qty|\alpha^X_n| = 1$. Therefore, a natural choice for the terms $H_{X,l}^{[n]}$ distributes the target parameters according to the weights:
\begin{equation}
    H_{X,l}^{[n]}(s) = \alpha^X_n \qty(f_l(s) - \mathrm{sign}(\alpha^X_n) \xi^X_l).
\end{equation}
For $X=A,D$, the sign function is redundant since all weights $\alpha^A_n, \alpha^D_n > 0$ are positive for all ply-indices $n$.

Accordingly, the full loss function Hamiltonian is formulated to be quadratic in local operators $\hat{H}_{X,l}^{[n]}$ on the individual subsystems $n$:
\begin{equation} \label{eq:loss_ham_2}
    \hat{H} = \sum_{X,l}\qty(\sum_{n=1}^N \hat{H}_{X,l}^{[n]})^2,
\end{equation}
where the local operators defined through the local functions $H_{X,l}^{[n]}(s)$ as:
\begin{equation} \label{eq:ham_loss_loc}
    \hat{H}_{X,l}^{[n]} = \sum_{t=1}^d H_{X,l}^{[n]}(t) \dyad{t}_n.
\end{equation}
Here, $\dyad{t}_n$ denotes the outer product of state $\ket{t} \in \calH_n$ and satisfies $\qty(\dyad{t}_n)\kv{s} = \delta_{t,s_n} \kv{s}$.

\subsection{Enforcing manufacturing constraints} \label{sec:constraints}

In addition to optimizing for the target lamination parameters, the stacking sequences must also adhere to manufacturing constraints, such as the ones listed in \cref{sec:stackingretrieval}. Here, we examine how these constraints, generalized to larger classes of constraints, can be enforced. 

Some constraints can be directly incorporated into the representation of the stacking sequences. For instance, to enforce symmetry around the midplane in a laminate, $s_n = s_{N-n+1}$, it suffices to model half of the stack and then implicitly mirror this to the other half. Due to the symmetry of the weights, the $A$ and $D$ lamination parameters in \cref{eq:lp} can be adapted to this case, by only summing the upper half $n=N/2+1,...,N$ of the stack and doubling the result. For a more convenient expression, we can reassign the ply indices $n$ to start at the midplane by substituting $n \to n-N/2$, such that $z_n = n$. We also redefine the ply number $N$ to just count half of the stack $N/2\to N$.  The reformulated lamination parameters only differ from \cref{eq:lp}  by the definition of $z_n$ and the factor $4$ preceding the summation in the $D$-parameters:
\begin{align} \label{eq:lpsym}
    (v^A_1,v^A_2,v^A_3,v^A_4) &= \frac{1}{N} \sum_{n=1}^N (z_n - z_{n-1}) \qty(\cos(2\theta_n),\sin(2\theta_n),\cos(4\theta_n),\sin(4\theta_n)) \\
    (v^D_1,v^D_2,v^D_3,v^D_4) &= \frac{1}{N^3}\sum_{k=1}^N (z_n^3 - z_{n-1}^3) \qty(\cos(2\theta_n),\sin(2\theta_n),\cos(4\theta_n),\sin(4\theta_n)) \notag\\
    z_n &= n \notag
\end{align}
As before, the weights:
\begin{equation}
    \alpha^A_n = \frac{1}{N}, \qquad \alpha^D_n = \frac{n^3 - (n-1)^3}{N^3}
\end{equation}
sum to 1. The $B$-parameters vanish for symmetric laminates and can thus be neglected. 

Enforcing the remaining constraints $c:\mathcal{S}\to\{\mathrm{True},\mathrm{False}\}$ in \cref{sec:stackingretrieval} can be achieved by incorporating a penalty function $H_c : \mathcal{S} \to \mathbb{R}_{\geq 0}$ into the loss function:
\begin{equation}
    H_{\mathrm{total}}(\vec{s}) = H(\vec{s}) + H_c(\vec{s}),
\end{equation}
where $H_c(\vec{s}) > 0$ if the constraint is violated for stacking sequence $\vec{s}$, $c(\vec{s}) = \mathrm{False}$, and $H_\mathrm{penalty}(\vec{s}) = 0$ for valid states, $c(\vec{s}) = \mathrm{True}$.
This effectively turns the constraint optimization problem with loss function $H$ into an unconstrained optimization problem with loss function $H_{\mathrm{total}}$ \cite{Lucas2014,Hen2016,Glover2022}. Often, a scalar factor is multiplied to the penalty function $H_c$, in order to control the magnitude of the penalty. However, this factor can also be regarded as part of the definition of the penalty function $H_c$. Accordingly, we include tunable parameters in the definition of the following penalty functions that we define for the different constraints in section \cref{sec:stackingretrieval}.

As with the loss function $H$, the penalty function $H_c$ is also transformed into a linear operator $\hat{H}_c$:
\begin{equation}
    \hat{H}_c = \sum_{\vec{t} \in \calS} H_c(\vec{t}) \dyad{\vec{t}}
\end{equation}
for its quantum representation. Consequently, the total Hamiltonian takes the form
\begin{equation}
    \hat{H}_{\mathrm{total}} = \hat{H} + \hat{H}_c.
\end{equation}
As discussed for the Hamiltonian $\hat{H}$, the interactions between subsystems in  $\hat{H}_c$ significantly influence the performance of various quantum algorithms. In the following, we explore possible penalty terms for the remaining manufacturing constraints apart from the symmetry constraint in \cref{sec:stackingretrieval}. For this purpose, we examine general categories of constraints that encompass the listed constraints.

\subsubsection{The disorientation constraint as nearest-neighbor coupling} \label{sec:constrnn}

The simplest constraint listed in \cref{sec:stackingretrieval} is the disorientation constraint, which limits the ply angle difference between neighboring plies. 
In general, a constraint with nearest neighbor coupling
can be enforced with a penalty function $\eta: \{1,\dots,d\}^2 \to \R_{\geq 0}$ on each pair of neighboring subsystems $n$ and $n+1$:
\begin{equation}
    \eta(s_n,s_{n+1}) = \begin{cases}
\gamma, & \text{if $(s_n,s_{n+1})$ violates  the constraint,}\\
0, & \text{else,}
\end{cases}
\end{equation}
where the adjustable parameter $\gamma > 0$ is the added penalty for a single constraint violation. The function $\eta$ is used to define an operator:
\begin{equation}
    \hat{H}_\mathrm{nn}^{[n,{n+1}]} = \sum_{t,t' = 1}^d \eta(t,t') \dyad{t t'}_{n,n+1}.
\end{equation}
As before, the indices of $\dyad{t t'}_{n,n+1}$ signify the local action of the operator on subsystems $\mathcal{H}_n$ and $\mathcal{H}_{n+1}$. Subsequently, these operators are summed over all nearest-neighbor pairs to yield the total penalty:
\begin{equation}
    \hat{H}_\mathrm{nn} = \sum_{n=1}^{N-1} \hat{H}_\mathrm{nn}^{[n,{n+1}]}. \label{eq:hamnn}
\end{equation}
The penalty operator exhibits strong parallels with quantum systems that feature nearest-neighbor coupling, such as spin lattice systems, and extensive research has been conducted on these systems \cite{Kadowaki1998,Parkinson2010,Kogut1979,Abrams1997,Poulin2009,Eisert2010,Bian2010,Cervera2018,Pagano2020,Kim2023}.

\subsubsection{The contiguity constraint as \texorpdfstring{$k$}{k}-local coupling}

The contiguity constraint restricts the number of consecutive same-angle plies to be equal or less than some small integer $N_\mathrm{same}$. A suitable penalty function evaluates $N_\mathrm{same}+1$ neighboring plies, and adds a penalty, if all plies have the same angle. The contiguity constraint is therefore a special case of a $k$-local constraint with $k = N_\mathrm{same}+1$, that encompasses interactions in a neighborhood of $k$ subsystems in a linear chain. The treatment of nearest-neighbour interactions above can be extended to accommodate this case, by defining:
\begin{equation}
    \eta(s_n,s_{n+1},\dots,s_{n+k-1}) = \begin{cases}
\gamma, & \text{if $(s_n,s_{n+1},\dots,s_{n+k-1})$ violates the constraint,}\\
0, & \text{else,}
\end{cases} 
\end{equation}
on $k$ consecutive subsystems. As before, the parameter $\gamma$ controls the magnitude of the added penalty. The corresponding $k$-local operator is given by:
\begin{equation}
    \hat{H}_{k\text{-local}}^{[n,\dots,{n+k-1}]} = \sum_{t_1,\dots,t_k = 1}^d \eta(t_1,\dots,t_k) \dyad{t_1 \cdots t_k}_{n,\dots,n+k-1}.
\end{equation}
The total penalty takes the form:
\begin{equation} \label{eq:hamklocal}
    \hat{H}_{k\text{-local}} = \sum_{n=1}^{N-k+1} \hat{H}_{k\text{-local}}^{[n,\dots,{n+k-1}]}.
\end{equation}
As discussed before, algorithms such as QAOA (\cref{sec:qaoa}) or quantum phase estimation (QPE) \cite{Kitaev1995,Abrams1999} require the implementation of interactions as operations. These higher-order terms typically require decomposition into two-qubit operations, leading to a significant increase in the total number of operations. Similarly, quantum annealers, which inherently support only second-order interactions, necessitate the transformation of higher-order terms into quadratic forms \cite{Kadowaki1998,Kadowaki2002,Neven2008,Bian2010,Rieffel2015,Hauke2020}. This often involves the introduction of auxiliary qubits as additional variables \cite{Rosenberg1975,Anthony2017,Dattani2019,Boros2020}. Nonetheless, with advancements in quantum technology, future devices could be capable of managing these heightened operational demands, especially if $k$ remains within a manageable range. 

\subsubsection{Balanced laminates and ensuring the same ply count for two distinct states}

The constraint for balanced laminates requires that the ply angle sequence contains an equal number of plies with angle $+\theta$ and angle $-\theta$, provided $\theta$ is neither $0^\circ$ nor $90^\circ$. This represents a special case of constraints where two states, denoted with $t$ and $t'$, appear an equal number of times in the stacking sequence:
\begin{equation}
    \sum_{n=1}^N \delta_{s_n,t} = \sum_{n=1}^N \delta_{s_n,t'} .
\end{equation}
To implement such a penalty function, we can take the squared difference, similar to the loss function of the lamination parameters in \cref{eq:loss_func}:
\begin{equation}
    H_\mathrm{balanced} (\vec{s}) = \gamma \qty(\sum_{n=1}^N \delta_{s_n,t} - \sum_{n=1}^N \delta_{s_n,t'})^2 = \gamma \qty(\sum_{n=1}^N \qty(\delta_{s_n,t} - \delta_{s_n,t'}))^2 .
\end{equation}
As before, the magnitude of the penalty can be adjusted with the parameter $\gamma \in \R_{>0}$. 
In order to define the corresponding operator, we replace a Kronecker delta $\delta_{s_n, t}$ with the dyadic product $\dyad{t}_n$ on subsystem $\calH_n$, which turns the classical penalty function into an operator:
\begin{equation} \label{eq:hambalanced}
    \hat{H}_\mathrm{balanced} = \gamma \qty(\sum_{n=1}^N \qty(\dyad{t}_n - \dyad{t'}_n))^2 ,
\end{equation}
where $\dyad{t}_n - \dyad{t'}_n$ is a local operator acting on subsystem $n$. This leads to a structure analogous to the Hamiltonian in \cref{eq:loss_ham_2} for the lamination parameters, featuring interactions between each pair of subsystems. Consequently, the earlier considerations regarding quantum algorithms are also relevant here. 

\subsubsection{The 10\%-rule and constraints on the ply count for a given state} \label{sec:penalty10perc}

The last manufacturing constraint listed in \cref{sec:stackingretrieval} is the $10\%$-rule, which necessitates that certain ply angles, typically $0^\circ$,$90^\circ$ or $\pm 45^\circ$, make up at least $10\%$ of the plies. This constitutes a special case of constraints that, for some state $t$, require at least $N_t$ plies to be in state $t$. A suitable penalty function can be defined as:
\begin{equation} \label{eq:penalty10perc}
    H_{\text{$10\%$-rule}}(\vec{s}) =  \begin{cases}
    0, & \text{if } \sum_{n=1}^N \delta_{s_n,t} \geq N_t,\\
    p\qty(N_t - \sum_{n=1}^N \delta_{s_n,t}), & \text{else,}
\end{cases}
\end{equation}
where the function $p(x)$ characterizes the behavior for constraint violations. Examples include constant, $p(x)=\gamma$, linear $p(x)= \gamma x$ or power-law $p(x) = \gamma x^\alpha$ behavior for some $\alpha,\gamma >0$. The according operator is defined as:
\begin{equation} \label{eq:ham10percent}
    \hat{H}_{\text{$10\%$-rule}} = \sum_{\vec{t}\in\mathcal{S}} H_{\text{$10\%$-rule}}(\vec{t}) \dyad{\vec{t}}
\end{equation}
As before, decomposing the operator into local operators provides insight into dependencies between individual subsystems. As we show in the appendix \ref{sec:app_penalty10perc}, any penalty operator suitable for the $10\%$-rule contains terms of order $N-N_t+1$ or larger. For the $10\%$-rule specifically, $N_t = 0.1N$, the operator contains terms that include roughly $90\%$ of all subsystems.
These high-order interactions present a notable challenge for many algorithms. Further investigation is vital to determine if and how these interactions can be efficiently managed in quantum algorithms. 

%% file: IV_algos.tex
\section{Selected algorithms in the context SSR} \label{sec:algo}

In the previous section, we discussed the representation of the stacking sequence retrieval problem using a basis-state encoding, along with a Hamiltonian that encodes the loss function and penalties.  This approach is general and applicable to a broad range of algorithms. This is especially true for \textit{variational quantum algorithms (VQA)}  \cite{Peruzzo2014,Farhi2014,Cerezo2021,Endo2021,Bharti2022}  and quantum annealing \cite{Hauke2020,Kadowaki1998,Kadowaki2002,Neven2008,Bian2010,Rieffel2015}, which are anticipated to be among the first methods to find real-world applications on near-term quantum hardware. Especially VQA are a diverse family of algorithms with various state-space parametrization, measurement and optimization strategies \cite{Cerezo2021,Endo2021,Bharti2022}, with ongoing research rapidly extending the amount of available methods. Identifying the optimal strategy for stacking sequence retrieval and integrating novel approaches will be crucial for demonstrating the effectiveness of quantum computation in this area. Here, we restrict the discussion to basic standard algorithms in order to illustrate how the presented quantum representation can be utilized for quantum computation. Specifically, we selected two different VQA for which we conduct state-vector simulations: a hardware efficient approach and the standard \textit{quantum approximate optimization algorithm (QAOA) } \cite{Farhi2014, Zhou2020}. Additionally, we utilize a tensor-network algorithm, the \textit{density matrix renormalization group (DMRG)} algorithm  \cite{Schollwock2011,Chan2016,White1992,White1993,Nishino1996,Dukelsky1998,Verstraete2004}, which is a classical variational algorithm closely related to one of the VQAs under investigation.

What all these variational algorithms share is a common objective: to minimize the expectation value of the Hamiltonian: 
\begin{equation}
    \ket{\psi_\mathrm{opt}} = \argmin_{\ket{\psi} \in \calH} \ev{\hat{H}}{\psi}.
\end{equation}
The resulting state $\ket{\psi_\mathrm{opt}}$ represents either the basis state corresponding to the optimal solution or -- in cases of multiple optimal configurations with identical minimal loss -- a potential superposition of optimal basis states. Although this objective unifies variational algorithms, they are differentiated by the specific strategies used to traverse the state space and achieve minimization.

In our demonstrations, we employ the four commonly used angles:
\begin{align}
    \theta_1 &= 0^\circ, & \theta_2 &= +45^\circ, & \theta_3 &= 90^\circ, & \theta_4 &= -45^\circ.
\end{align}
Assuming a symmetric laminate, we utilize the corresponding lamination parameters as detailed in \cref{eq:lpsym}. This section explores the implications of various constraints outlined in \cref{sec:constraints} within the context of these algorithms. For our numerical demonstrations in \cref{sec:demo}, we include a disorientation constraint that limits the ply-angle difference between adjacent plies to $45^\circ$ or less. This constraint is breached when neighboring plies are $\theta_1$ and $\theta_3$, or $\theta_2$ and $\theta_4$, and is enforced via a penalty term as discussed in \cref{sec:constrnn}.

In the remainder of this section, we delve into algorithm-specific considerations. To conclude, we discuss the similarities between the chosen hardware-efficient approach and DMRG, in contrast to QAOA. The results of our numerical demonstrations are detailed in the following \cref{sec:demo}.

\subsection{Variational quantum algorithms} \label{sec:vqa}

\textit{Variational quantum algorithms (VQA)} employ parameterized quantum circuits to define a subspace of the Hilbert space $\calH$ using real, continuous parameters \cite{Cerezo2021,Endo2021}. By configuring a quantum circuit with $N_p$ parameters $\vec{p} \in \R^{N_p}$, we can prepare the state $\ket{\psi(\vec{p})} \in \calH$ on a quantum computer. This setup enables the repeated preparation and measurement of this state to estimate the Hamiltonian's expectation value, $\ev*{\hat{H}}_\psi (\vec{p}) = \ev{\hat{H}}{\psi(\vec{p})}$. The minimization itself is carried out by a classical optimizer, which iteratively selects parameters $\vec{p}$ and executes calls to the function $\ev*{\hat{H}}_\psi (\vec{p})$ based on its specific strategy. Leveraging limited quantum resources for essential computations while relying on classical systems for optimization makes VQAs particularly suitable for near-term quantum devices.

Most contemporary quantum computers are based on qubits, which are characterized by two basis states $\ket{0_q}$ and $\ket{1_q}$. To distinguish the qubit states from the ply states, we apply a subscript $q$ to denote qubit states. The basis states $\ket{\vec{s}} \in \calH$ therefore need to be represented in terms of qubit states. In the case of the above mentioned 4 ply-angles, a subspace $\calH_n$ can be implemented as 2 qubits using a binary encoding, which we specifically choose to be:
\begin{align}
    \ket{1} &\equiv \ket{00_q} & \ket{2} &\equiv \ket{01_q} & \ket{3} &\equiv \ket{11_q} & \ket{4} &\equiv \ket{10_q}. \label{eq:encoding}
\end{align}
In this ply-state encoding, flipping one of the qubits corresponds to a ply-angle change of $+45^\circ$ or $-45^\circ$, while changing both qubits results in a $90^\circ$ rotation.

For a general non-diagonal Hamiltonian, the expectation value is typically determined by expressing the Hamiltonian using Pauli operators $\hat{X}$, $\hat{Y}$, and $\hat{Z}$ applied to the qubits:
\begin{align}
    \hat{X} &= \dyad{0_q}{1_q} + \dyad{1_q}{0_q} & Y &= i \ (\dyad{0_q}{1_q} - \dyad{1_q}{0_q}) & Z &= \dyad{0_q} - \dyad{1_q}.
\end{align}
The expectation value for individual terms is measured using appropriate basis rotations. Due to linearity, the total expectation value equates to the sum of these individual contributions. In our specific scenario, however, the Hamiltonian is diagonal in the basis of stacking sequences and, by extension, in the qubit basis. This configuration allows for straightforward repeated measurements of a state and aggregation of loss function values from individual outcomes, divided by the total number of counts. Although this simplifies the measurement process, formulating the Hamiltonian in terms of Pauli operators remains a necessity for specific algorithms, as demonstrated for QAOA in \cref{sec:qaoa}. 

As mentioned earlier, VQAs exhibit significant variability in state preparation circuits -- ranging from problem-inspired \cite{Farhi2014,Hadfield2019,Peruzzo2014,McClean2016,Lee2019} to hardware efficient \cite{Kandala2017,Moll2018} --but also in state measurements and optimization strategies. Recent developments includes adaptive approaches to enhance the performance of the algorithm \cite{Grimsley2019,Zhu2022,Bravyi2021,Amaro2022, Patel2024,Garcia2019}, and identifying VQAs that are optimally suited for problems like stacking sequence retrieval will necessitate extensive research and will likely hinge on ongoing advancements in this dynamic field. In this work, we focus on demonstrating our methodology using two specific standard VQAs: a hardware-efficient approach and the standard QAOA.

\subsubsection{A hardware-efficient approach} \label{sec:hwe}

\begin{figure}[t]
    \centering
    \includegraphics[scale=1.0]{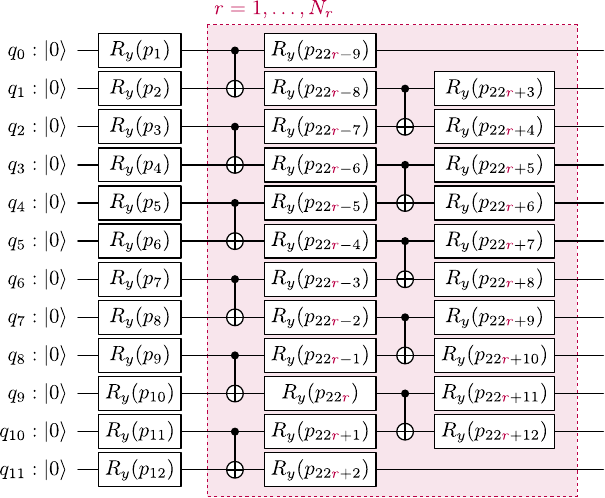}
    \caption{Parameterized quantum circuit for hardware-efficient approach}
    \label{fig:hwe_pqc}
\end{figure}

Among variational quantum algorithms, hardware-efficient approaches are distinguished by their utilization of short-depth quantum circuits. These circuits efficiently utilize available gates within a specific quantum computing architecture to parameterize a substantial portion of the state space. In this study, we employ a state preparation circuit, as illustrated in \cref{fig:hwe_pqc}, which incorporates $\hat{R}_y(p)$ single-qubit rotations to facilitate dependency on the continuous parameters $p$, and CNOT gates to generate entanglement among the qubits. Typically, in hardware-efficient approaches, the state preparation circuit features a block of gates that is repeated a certain number of times. Adjusting the number of repetitions modulates the number of parameters, enabling exploration of a larger subspace at the cost of increasing the complexity of optimization. A particular feature of our selected state preparation circuit is the exclusive connection of neighboring qubits by the CNOT gates. This particular state preparation circuit has previously been employed, for instance in \cite{Amaro2022}.

A significant obstacle in hardware-efficient VQA is the optimization of a large number of parameters \cite{Bittel2021} and several strategies have been developed to tackle this issue. These include techniques that optimize only a subset of parameters at a time, such as optimizing different segments of the parameterized circuit sequentially \cite{Lyu2020,Skolik2021,Liu2022b}. In our approach, we adopt an optimization strategy that sequentially optimizes two neighboring qubits at a time, starting either from the qubits corresponding to plies $n=1$ or $n=N$. This sequence progresses one qubit at a time, culminating after $N-1$ pairwise optimizations. The process involves multiple sweeps of local optimizations across the entire stack until the optimization converges. This method bears notable similarities to the DMRG algorithm, which we discuss subsequent to the QAOA.

\subsubsection{QAOA} \label{sec:qaoa}

The most recognized variational quantum algorithm for combinatorial optimization is the \textit{quantum approximate optimization algorithm (QAOA)} \cite{Farhi2014,Farhi2016}, which employs two distinct types of operators in its state preparation circuit. The problem unitary is defined by the exponential of the Hamiltonian:
\begin{equation} \label{eq:problem_u}
    U_C (\gamma) = \exp(-i \gamma \hat{H}),
\end{equation}
and the mixer unitary is articulated through the exponential of $\hat{X}$ operators applied to all $N_q = 2N$ qubits:
\begin{equation}
    U_M(\beta) = \exp(-i \beta \sum_{j=1}^{N_q} \hat{X}_j),
\end{equation}
where $\hat{X}_j$ represents the operator $\hat{X}$ on qubit $j$. Each operator introduces a single parameter -- $\beta \in \R$ or $\gamma \in \R$ -- into the circuit. The entire circuit begins by applying single-qubit rotations to each qubit to prepare the state $\ket{+}^{\otimes {N_q}}$, where $\ket{+} = \frac{1}{\sqrt{2}} (\ket{0_q} + \ket{1_q})$. Subsequently, a defined number of $N_r$ repetitions of the problem and mixer unitaries are alternately executed on this initial state:
\begin{equation}
    \ket{\psi(\vec{\beta},\vec{\gamma})} = U_M(\beta_{N_r}) U_C(\gamma_{N_r}) \dots U_M(\beta_{1}) U_C(\gamma_{1}) \ket{+}^{\otimes {N_q}}. 
\end{equation}
Thus, the complete quantum circuit that prepares $\ket{\psi(\vec{\beta},\vec{\gamma})}$ comprises $2 N_r$ continuous parameters that need optimization.

To implement the state preparation circuit on a quantum computer, the problem and mixer unitaries must be converted into a sequence of available gates. As common practice, we focus on the single-qubits rotations $\hat{R}_x$, $\hat{R}_y$ and $\hat{R}_z$, and the CNOT gate. Given that shorter circuits are quicker to evaluate and less susceptible to errors, the total number of gates in the decomposition significantly influences the performance of the algorithm. The mixer unitary, which is always the same for standard QAOA, is typically decomposed into a layer of parallel single-qubit rotations and therefore only has a minor impact on the circuit depth. 

Conversely, the specific form of the Hamiltonian significantly affects the gate count required for the problem unitary. Initially, we analyze the loss function Hamiltonian in \cref{eq:loss_ham_2}, which is expanded to take the form:
\begin{equation}
    \hat{H} = \sum_{X,l}\sum_{n,n'=1}^N \hat{H}_{X,l}^{[n]} \hat{H}_{X,l}^{[n']}.
\end{equation}

Furthermore, we can substitute the definition of the local matrices in \cref{eq:ham_loss_loc} to obtain:
\begin{equation} \label{eq:ham4qaoa_expand}
\hat{H} = \sum_{X,l} \sum_{n,n'=1}^N \sum_{t,t'=1}^d \hat{H}_{X,l}(t_n) \hat{H}_{X,l}(t'_{n'})  \dyad{tt'}_{nn'}.
\end{equation}
The operators $\dyad{tt'}_{nn'} = \dyad{t}_n \dyad{t'}_{n'}$ act on subsystems $n$ and $n'$. Following the encoding in \cref{eq:encoding}, we express an operator $\dyad{s}$ on a single subsystem with qubit operators as follows:
\begin{align}
\dyad{1} \equiv \dyad{00_q} &= \frac{1}{4} \qty(\hat{I} \otimes \hat{I} + \hat{I} \otimes \hat{Z} + \hat{Z} \otimes \hat{Z} + \hat{Z} \otimes \hat{I}), \\
\dyad{2} \equiv \dyad{01_q} &= \frac{1}{4} \qty(\hat{I} \otimes \hat{I} - \hat{I} \otimes \hat{Z} - \hat{Z} \otimes \hat{Z} + \hat{Z} \otimes \hat{I}) , \notag \\
\dyad{3} \equiv \dyad{11_q} &= \frac{1}{4} \qty(\hat{I} \otimes \hat{I} - \hat{I} \otimes \hat{Z} + \hat{Z} \otimes \hat{Z} - \hat{Z} \otimes \hat{I}) , \notag \\
\dyad{4} \equiv \dyad{10_q} &= \frac{1}{4} \qty(\hat{I} \otimes \hat{I} + \hat{I} \otimes \hat{Z} - \hat{Z} \otimes \hat{Z} - \hat{Z} \otimes \hat{I}) . \notag
\end{align}
Consequently, an operator $\dyad{tt'}_{nn'}$ contains terms with up to $d=4$ $\hat{Z}$-operators. After this substitution, all terms in the Hamiltonian with the same operators can be collected. 

Given that terms of the Hamiltonian $\hat{H}$ only contain $\hat{Z}$-operators, they all commute. The exponential function in the problem unitary $U_C$ in \cref{eq:problem_u} can therefore be separated into the exponential function of the individual terms. The number of terms therefore has a direct impact on the depth of the decomposed circuit. Furthermore, the exponential function of a term comprising only identity operators and a number of $N_Z$ Pauli-$\hat{Z}$ operators can be implemented with $2 N_Z$ CNOT-gates and a single-qubit $\hat{R}_z$ rotation \cite{Nielsen2012}. 
It is therefore favorable to only have low-order terms. Due to its construction and the chosen qubit-encoding, the loss function Hamiltonian only contains terms up to fourth order. The expansion also includes a scalar term, only consisting of identity operators, which merely adds a phase to the exponential and can thus be ignored.

We can calculate the number of possible terms by examining the different scenarios: There are $2N$ ways to place a single $\hat{Z}$ operator. For second-order terms, there are $N$ ways to place $\hat{Z}$ on a pair of qubits corresponding to one ply and $4N(N-1)$ ways to place two $\hat{Z}$ operators on different plies. For third-order terms, there are $2N(N-1)$ ways to place a pair of $\hat{Z}$ operators on one ply and a single one on another ply. Lastly, there are $N(N-1)$ possible fourth-order terms resulting from placing two pairs on two different plies. Consequently, the unitary $U_C$ contains at most $2N$ terms with one, $4N^2-3N$ terms with two, $2N(N-1)$ terms with three, and $N(N-1)$ terms with four $\hat{Z}$ operators. The quadratic nature of the Hamiltonian results in an $O(N^2)$ behavior in both the number of terms and the required number of two-qubit gates in the problem unitary $U_C$. Given that this circuit is executed $N_r$ times, the complexity can significantly increase with a larger number of plies $N$, potentially exceeding the capabilities of near-future quantum devices. Nevertheless, the quadratic scaling and the lower order of the terms are generally viewed as favorable when compared to the potential complexity of terms in a general Hamiltonian.

We can similarly analyze the penalty operators described in \cref{sec:constraints}. The nearest-neighbor constraint, detailed in \cref{eq:hamnn}, comprises solely second-order terms. Once the qubit encoding is applied, these terms yield at most four $\hat{Z}$ operators. Since these operators are already incorporated into the loss function Hamiltonian, the integration of the penalty terms and subsequent simplification of the Hamiltonian do not increase the number of terms or their order. Consequently, the circuit depth remains unchanged upon adding the penalty term. Similarly, the penalty for balanced laminates, as specified in \cref{eq:hambalanced}, is quadratic and can be added to the Hamiltonian without increasing the gate count.

The situation changes when considering the penalty operators for the $k$-local constraint in \cref{eq:hamklocal}. These terms involve operators on $k$ consecutive plies, and after encoding, the number of $\hat{Z}$ operators can escalate to $2k$, necessitating up to $4k$ two-qubit gates per term. However, since the operators in a term are restricted to consecutive plies, the total number of possible terms scales as $O(N)$, assuming $k$ remains constant. Thus, if $k$ is not excessively large, the gate count required to implement this operator might still be manageable on near-term quantum computers.

As previously discussed in \cref{sec:penalty10perc}, any expansion of the penalty operator for the $10\%$-rule, as delineated in \cref{eq:ham10percent}, introduces high-order terms that scale with the number of plies $N$. Implementing the exponential function of these high-order terms presents a significant challenge to enforcing the $10\%$-rule in QAOA using a penalty term.

In our demonstrations in section \cref{sec:demo}, we focus exclusively on implementing the disorientation constraint as a nearest-neighbor constraint.

\subsection{DMRG}

The hardware-efficient approach, along with the selected optimization procedure, shares many similarities with a specific tensor network method known as the \textit{density matrix renormalization group (DMRG)} algorithm \cite{Schollwock2011,Chan2016,White1992,White1993,Nishino1996,Dukelsky1998,Verstraete2004}. A principal challenge in simulating quantum-based calculations on classical computers is the exponentially large state space, making it impractical to represent a state by storing all its components. Tensor networks address this challenge by enabling the representation of high-order tensors through a network of interconnected low-order tensors \cite{Orus2014,Orus2019,Bridgeman2017,Biamonte2017,Cirac2021,Evenbly2022,Oseledets2009,Oseledets2011}. Although these methods have limitations and are unlikely to efficiently simulate full-scale fault-tolerant quantum computations, they have proven highly effective in studying many-body quantum mechanics and simulating quantum circuits. Remarkably, some quantum-computing experiments that claimed to demonstrate quantum advantage were later classically simulated with tensor network methods (\cite{Arute2019} and \cite{Gray2021,Pan2021}, \cite{Kim2023} and \cite{Tindall2024,Begusic2024,Parta2024}).

The DMRG algorithm is a specialized tensor network method designed to identify the ground state of one-dimensional many-body quantum systems. It employs specific tensor network structures -- \textit{matrix product states (MPS)} and \textit{matrix product operators (MPO)} -- to represent quantum states and operators, respectively. The optimization procedure of the DMRG algorithm mirrors that of our hardware-efficient approach described in \cref{sec:hwe}. Specifically, much like only adjacent qubits are interconnected by CNOT gates in our approach, the MPS configuration connects local tensors across neighboring subsystems. Moreover, the optimization in DMRG involves local optimizations on neighboring subsystem pairs, which is akin to the localized pairwise optimization of adjacent qubits used in our hardware-efficient strategy. In fact, and as we discuss later on, the hardware-efficient circuit can easily be converted into an MPS. A key distinction from the quantum algorithms is that MPS and MPO representations are defined directly on the ply-subsystems $\calH_n$, eliminating the need for qubit decomposition. Thus, optimizations are conducted on neighboring subsystems rather than qubits.

The main challenge of applying the DMRG algorithm to the SSR problem is finding an MPO representation of the loss function Hamiltonian and the penalty operators. 
We begin our discussion by developing an MPO representation for the loss function, followed by an examination of the various penalty operators in \cref{sec:constraints}.

\subsubsection{An MPO representation of the loss function Hamiltonian} \label{sec:mpoloss}

In this section, we construct an MPO representation of the loss function Hamiltonian $\hat{H}$ in \cref{eq:loss_ham_2}. As we discuss further below, a summation of matrix product operators can straightforwardly implemented in MPO form. Thus, we initially concentrate on a single term $\hat{H}_{X,l}$, corresponding to the lamination parameter $v^X_l$:
\begin{equation} \label{eq:ham_xl}
\hat{H}=\sum_{X,l} \hat{H}_{X,l}, \qquad \hat{H}_{X,l} = \qty(\sum_{n=1}^N \hat{H}^{[n]}_{X,l})^2.
\end{equation}
For clarity, we define the matrix elements for states $\ket{\vec{s}}, \ket{\vec{s'}} \in \calH$ of the diagonal operator $\hat{H}_{X,l}$ as:
\begin{equation} \label{eq:ev_hamxl}
 h_{\vec{s},\vec{s}'} := \mel{\vec{s}}{\hat{H}_{X,l}}{\vec{s}'} = \delta_{\vec{s},\vec{s}'} h_{\vec{s}}, \qquad h_{\vec{s}} = \mel{\vec{s}}{\hat{H}}{\vec{s}},
 \end{equation}
implying:
\begin{equation}
    \hat{H}_{X,l} = \sum_{\vec{t},\vec{t}' \in \calS} h_{\vec{t},\vec{t}'} \dyad{\vec{t}}{\vec{t}'} = \sum_{\vec{t} \in \calS} h_{\vec{t}} \dyad{\vec{t}}.
\end{equation}
Similarly, we denote the eigenvalues of the local operators $\hat{H}^{[n]}_{X,l}$ as:
\begin{equation}
    h^{[n]}_{s_n} = \ev{\hat{H}^{[n]}_{X,l}}{s_n} = H^{[n]}_{X,l} (s_n),
\end{equation}
where, according to the definition of the loss function, we have:
\begin{equation} \label{eq:ev_eq}
    h_{\vec{s}} = \qty(\sum_{n=1} h_{s_n}^{[n]} )^2 .
\end{equation}

To construct the MPO representation, we aim to define a tensor $W^{[1]}$ sized $d \times d \times b_1$, a tensor $W^{[N]}$ sized $d \times d \times b_{N-1}$, and tensors $W^{[n]}$ for $1 < n < N$ sized $d \times d \times b_{n-1} \times b_{n}$, such that:
\begin{equation} \label{eq:loss_mpo}
    h_{\vec{s},\vec{s}'} = W^{[1]}_{s_1 s_1'} W^{[2]}_{s_2 s_2'} \cdots W^{[N-1]}_{s_{N-1} s_{N-1}'} W^{[N]}_{s_N s_N'},
\end{equation}
where the bond dimensions $b_1,\cdots,b_{N-1}$ are parameters yet to be determined based on the specific MPO representation required. Note, that for specific $\vec{s},\vec{s}'\in\calS$, the matrices $W^{[n]}_{s_n,s'_n}$ are multiplied with matrix multiplication, with the ones for $n=1$ and $n=N$ taken as a row and column vector respectively. Given that the Hamiltonian is diagonal, the tensors $W^{[n]}$ are also assumed diagonal in the indices $s_n$ and $s_n'$:
\begin{equation}
    W^{[n]}_{s_n s_n'} = \delta_{s_n s_n'} W^{[n]}_{s_n}, \qquad W^{[n]}_{s_n} := W^{[n]}_{s_n s_n}.
\end{equation}
Thus, to reflect \cref{eq:ev_eq}, matrices $W^{[n]}_{s_n}$ must be defined such that:
\begin{equation}
    \qty(\sum_{n=1} h_{s_n}^{[n]} )^2 = W^{[1]}_{s_1} W^{[2]}_{s_2} \cdots W^{[N]}_{s_N}, 
\end{equation}
and the entire MPO representation for the matrix elements $h_{\vec{s},\vec{s}'}$ can be realized by multiplying a Kronecker delta $\delta_{s_n,s_n'}$ with the resulting matrices $W^{[n]}_{s_n}$.

Our approach decomposes $h_{\vec{s}}$ using a matrix whose repeated multiplication yields $h_{\vec{s}}$ as the top-right element. Specifically, we define:
\begin{equation}
    M(x) = \bmat{1 & \sqrt{2} x & x^2 \\ 0 & 1 & \sqrt{2} x \\ 0 & 0 & 1 },
\end{equation}
which follows the multiplication rule:
\begin{equation}
    M(x) M(y) = M(x+y).
\end{equation}
By induction, multiplying $N$ of these matrices, each associated with the eigenvalues $h_{s_n}^{[n]}$, results in:
\begin{equation}
    \prod_{n=1}^N M(h^{[n]}_{s_n}) = M\qty(\sum_n h^{[n]}_{s_n}) = \bmat{1 & \sqrt{2} \qty(\sum_n h^{[n]}_{s_n}) & \qty(\sum_n h^{[n]}_{s_n})^2 \\ 0 & 1 & \sqrt{2} \qty(\sum_n h^{[n]}_{s_n}) \\ 0 & 0 & 1 } .
\end{equation}
An MPO representation of $\hat{H}_{X,l}$ as in \cref{eq:loss_mpo} can thus be obtained by breaking it down into these matrices:
\begin{equation}
    W^{[n]}_{s_n s'_n} = \delta_{s_n s'_n} M(h^{[n]}_{s_n})
\end{equation}
for $1 < n < N$, with the first row and third column of the first and last matrix selected to yield $\left(\sum_n h^{[n]}_{s_n}\right)^2$:
\begin{align}
    W^{[1]}_{s_1 s'_1} &= \bmat{1 & 0 & 0} \delta_{s_1 s'_1} M(h^{[1]}_{s_1}),\\
    W^{[N]}_{s_N s'_N} &= \delta_{s_N s'_N}M(h^{[N]}_{s_N})\bmat{0 \\ 0 \\ 1}. \notag
\end{align}

The total loss function Hamiltonian in \cref{eq:loss_ham_2} compiles $\hat{H}_{X,l}$ for all lamination parameters $X=A,B,D$ and $l=1,2,3,4$, excluding $X=B$ for symmetric laminates. This sum of matrix product operators (MPOs) can theoretically be realized by stacking the matrices $W^{[n]}_{s_n s'_n}$ for each $n$ into block-diagonal matrices. However, thanks to the linearity of the expectation value $\ev{\hat{H}}{\psi}$, the DMRG method can efficiently iterate over each term individually by calculating an environment tensor for each term and subsequently summing these tensors element-wise to construct the overall environment tensor.


\subsubsection{An MPO representation for nearest-neighbor constraints} \label{sec:mponn}

As discussed in \cref{sec:constraints}, we can implement nearest-neighbor constraints as penalty functions, which can be expressed as operators and added to the Hamiltonian. 
The sum of the operators in \cref{eq:hamnn} effectively counts the constraint violations. There, we defined the added penalty for one constraint violation to be
\begin{equation}
    \eta(s_n, s_{n+1}) = \begin{cases}
\gamma, & \text{if $(s_n, s_{n+1})$ violates constraint,}\\
0, & \text{else,}
\end{cases}
\end{equation}
where $\gamma > 0$ is the penalty for a single constraint violation.
Similar to other terms in the Hamiltonian, the penalty function needs to be represented as an MPO. As before, we construct a matrix that calculates the desired result, which is the count of constraint violations, in the top-right corner. For this purpose, we represent the function $\eta(s_n,s_{n+1})$ as the dot product of two vectors $\vec{p}(s_n)$ and $\vec{q}(s_{n+1})$, both of dimension $d$:
\begin{equation}
    \eta(s_n,s_{n+1}) = \vec{p}(s_{n}) \cdot \vec{q}(s_{n+1}) = \sum_{k = 1}^d p_k(s_n) q_k(s_{n+1}). 
\end{equation}
We achieve this by using $\vec{p}(s_{n})$ as an indicator for the state $s_n$, and $\vec{q}(s_{n+1})$ to indicate which neighboring ply of state $s_{n}$ leads to a constraint violation:
\begin{align} \label{eq:defpq}
    p_k(s_n) &= \delta_{k,s_n}, &
    q_k(s_{n+1}) &= \eta(k,s_{n+1}).
\end{align}
If we define a matrix $M(s)$ based on $\vec{p}$ and $\vec{q}$ as:
\begin{equation} \label{eq:matm}
    M(s) = \left[\begin{array}{ccccc}
    1 & p_1(s) & \cdots & p_d(s) & 0\\
    0 & 0 & \cdots & 0 & q_1(s) \\
    \vdots & \vdots & \ddots & \vdots & \vdots\\
    0 & 0 & \cdots & 0 & q_d(s) \\
    0 & 0 & \cdots & 0 & 1
\end{array}\right],
\end{equation}
it can be shown by induction that the product of these matrices yields the count of constraint violations in the top-right corner of the resulting matrix:
\begin{equation}
    \bmat{1 & 0 & 0} \qty(\prod_{n=1}^N M(s_n)) \bmat{0 \\ 0 \\ 1} =  \sum_{n=1}^{N-1} \eta(s_n,s_{n+1}).  \label{eq:constraintmpo}
\end{equation}
This matrix form is similar to the treatment of operators in \cite{Parker2020,McCulloch2007}. To retrieve the penalty in MPO form, we can add Kronecker deltas analogously to before.

In the particular case of a disorientation constraint of $45^\circ$, the constraint is violated when neighboring plies are  $\theta_1$ and $\theta_3$, or $\theta_2$ and $\theta_4$. In this case, we can define the vectors  $\vec{p}(s)$ and $\vec{q}(s)$ for the MPO representation of the penalty as:
\begin{align}
    p_k(s) &= \delta_{k,s}, &
    q_k(s) &= \gamma (\delta_{k,s+2} + \delta_{k,s-2}),
\end{align}
where $\gamma$ represents the penalty added for a single constraint violation.

\subsubsection{An MPO representation for the contiguity constraint} \label{sec:mpocontiguity}

For the contiguity constraint, we extend the approach for nearest-neighbor constraints to $k$-local constraints. This extension involves modifying the penalty function, which is now defined for a sequence of $k$ subsystems. Specifically, the penalty function $\eta(s_n, s_{n+1}, \dots, s_{n+k-1})$ assigns a penalty of $\gamma$ if the sequence $(s_n, s_{n+1}, \dots, s_{n+k-1})$ violates the constraint, and 0 otherwise. To represent this penalty function, we seek $k$ vectors $\vec{p}^{(1)}(s),\dots,\vec{p}^{(k)}(s)\in\C^{\tilde{d}}$ for each ply angle $s\in\{1,\dots,d\}$ such that:
\begin{equation}
     \eta(s_n,s_{n+1},\dots,s_{n+k-1}) = \sum_{j=1}^{\tilde{d}} p^{(1)}_j (s_n) \, p^{(2)}_j (s_{n+1}) \cdots p^{(k)}_j (s_{n+k-1}).
\end{equation}
Here, $d$ is the number of allowed ply angles, and $\tilde{d}$, a positive integer, denotes the size of these vectors. For the specific case of the contiguity constraint, which penalizes $k$ consecutive plies with the same ply-angle, we find an effective representation with $\tilde{d} = d$. The vectors are defined as:
\begin{equation}
    p^{(1)}_j(s) = \cdots = p^{(k-1)}_j(s) =  \delta_{js}, \qquad p^{(k)}_j(s) =  \gamma \delta_{js} , \label{eq:penaltycontiguity}
\end{equation}
where we chose $\gamma$ to included $p^{(k)}_j(s)$. Other possibilities include multiplying $\gamma$ to another vector or distribute it by multiplying $\gamma^{\frac{1}{k}}$ to each of the vector.

We extend the definition of the matrix $M(s)$, as outlined in eq. (\ref{eq:matm}), to adapt it for the $k$-local scenario. This adaptation involves constructing a square matrix of dimension $\tilde{d}(k-1)+2$. The vectors $\vec{p}^{(1)}(s)$ and $\vec{p}^{(k)}(s)$ are incorporated in the first row and last column respectively, similar to eq. (\ref{eq:matm}). However, to cater to the $k$-local aspect, we integrate additional columns and rows. These new elements form a diagonal matrix $\mathrm{diag}\qty(\vec{p}^{(2)}(s),\cdots,\vec{p}^{(k-1)}(s))$ of dimension $\tilde{d}(k-2)$, with the elements of vectors $\vec{p}^{(2)}(s)$ to $\vec{p}^{(k-1)}(s)$ listed sequentially along the diagonal. The resulting matrix $M(s)$ for the $k$-local case can thus be represented in block-matrix form as:
\begin{equation}
    M(s) = \left[\begin{array}{cccc}
      1 & \qty(\vec{p}^{(1)}(s))^{\mathrm{T}} & \mathbf{0} & 0 \\
     \mathbf{0} & \mathbf{0} & \mathrm{diag}\qty(\vec{p}^{(2)}(s),\cdots,\vec{p}^{(k-1)}(s)) & \mathbf{0} \\ 
    \mathbf{0}  & \mathbf{0} &\mathbf{0} & \vec{p}^{(k)}(s) \\
    \mathbf{0}  & \mathbf{0} & \mathbf{0} & 1
  \end{array}\right].
\end{equation}

In this matrix, $\qty(\vec{p}^{(1)}(s))^{\mathrm{T}}$ is a row vector and $\vec{p}^{(k)}(s)$ is a column vector, both of size $\tilde{d}$. The matrices $\mathbf{0}$ contain only zeros and are sized appropriately. Single-element matrices containing 0 or 1 are denoted in regular font weight. As in section \ref{sec:mponn}, a product of matrices $M(s_1)\cdots M(s_N)$ counts the penalties for constraint violations in the top right corner:
\begin{equation}
    \begin{bmatrix}
    1 & 0 & \cdots & 0
\end{bmatrix} M(s_1)\cdots M(s_N) \begin{bmatrix}
    0 \\ \vdots \\ 0 \\ 1
\end{bmatrix} = \sum_{n=1}^{N-k+1} \eta(s_n,\dots,s_{n+k-1}), \label{eq:mpoklocal}
\end{equation}
where the vectors surrounding the product select for the first row and last column. Following, the element-wise (or Hadamard) product of two vectors, $\vec{a}$ and $\vec{b}$, is denoted by $\vec{a}\odot\vec{b}$, and is defined as:
\begin{equation}
    (\vec{a}\odot \vec{b})_j = a_j b_j.
\end{equation}
To illustrate this, consider the action of a single matrix $M(s)$ on a vector. For an arbitrary complex number $a\in \C$ and vectors $\vec{b}^{(j)}\in\C^{\tilde{d}}$ for $j=1,\dots,k-1$, the matrix operation is given by:
\begin{equation}
    M(s) \begin{bmatrix}
    a \\ \vec{b}^{(1)} \\ \vec{b}^{(2)} \\ \vdots \\ \vec{b}^{(k-2)} \\ \vec{b}^{(k-1)}\\ 1
\end{bmatrix} = \begin{bmatrix}
    a + \vec{p}^{(1)}(s) \cdot \vec{b}^{(1)}  \\ \vec{p}^{(2)}(s)  \odot  \vec{b}^{(2)} \\ \vec{p}^{(3)}(s) \odot \vec{b}^{(3)}  \\ \vdots \\  \vec{p}^{(k-1)}(s) \odot \vec{b}^{(k-1)} \\ \vec{p}^{(k)}(s) \\ 1
\end{bmatrix}.
\end{equation}
In this operation, each vector $\vec{b}^{(j)}$ is elevated one position and then multiplied element-wise with the corresponding vector $\vec{p}^{(j)}(s)$. The top vector $\vec{b}^{(1)}$ undergoes a dot product with $\vec{p}^{(1)}(s)$, which is added to the scalar $a$. The vector $\vec{p}^{(k)}(s)$ is inserted into the bottom slot that becomes vacant. Consequently, when the matrices in equation (\ref{eq:mpoklocal}) are multiplied from right to left, we observe the sequential accumulation and elevation of the vectors $\vec{p}^{(k)}$. This is depicted in the following process:
\begin{align}
&\begin{bmatrix}
    0 \\ \vdots \\ 0 \\ 1
\end{bmatrix} \quad \overset{M(s_N)\cdot }{\longrightarrow} \quad \begin{bmatrix}
    0  \\ \mathbf{0}_{\tilde{d}(k-2)} \\ \vec{p}^{(k)}(s_N) \\ 1
\end{bmatrix} \quad \overset{M(s_{N-1})\cdot }{\longrightarrow} \quad \begin{bmatrix}
    0  \\ \mathbf{0}_{\tilde{d}(k-3)} \\ \vec{p}^{(k-1)}(s_{N-1}) \odot \vec{p}^{(k)}(s_N) \\ \vec{p}^{(k)}(s_{N-1}) \\ 1
\end{bmatrix} \quad \overset{M(s_{N-2})\cdot }{\longrightarrow} \quad \cdots \\
&\overset{M(s_{N-k+2})\cdot }{\longrightarrow} \quad \begin{bmatrix}
    0  \\ \vec{p}^{(2)}(s_{N-k+2}) \odot \cdots \odot \vec{p}^{(k)}(s_N) \\ \vdots \\ \vec{p}^{(k-1)}(s_{N-k+2}) \odot \vec{p}^{(k)}(s_{N-k+3}) \\ \vec{p}^{(k)}(s_{N-k+2}) \\ 1
\end{bmatrix} \quad \overset{M(s_{N-k+1})\cdot }{\longrightarrow} \quad \begin{bmatrix}
    \eta(s_{N-k+1},\dots,s_{N}) \\ \vec{p}^{(2)}(s_{N-k+1}) \odot \cdots \odot \vec{p}^{(k)}(s_{N-1}) \\ \vdots \\ \vec{p}^{(k-1)}(s_{N-k+1}) \odot \vec{p}^{(k)}(s_{N-k+2}) \\ \vec{p}^{(k)}(s_{N-k+1}) \\ 1
\end{bmatrix} . \notag
\end{align}
In this sequence, the suffixes of the matrices $\mathbf{0}$ indicate their size. Upon integrating the matrix $M(s_{N-k+1})$, the accurate penalty for the last $k$ subsystems, from $N-k+1$ to $N$, is added to the first row. The elements below are positioned to include the correct penalty for the subsystems from $N-k$ to $N-1$ when the subsequent matrix $M(s_{N-k})$ is integrated into the product. By continuing this process for all matrices and then selecting the first row, the correct penalty for the entire stacking sequence $\vec{s}$ is computed, validating equation (\ref{eq:mpoklocal}).

The matrices resulting from this formulation, each having a dimension of $\tilde{d} (k-1) + 2$, are notably larger compared to the other MPOs we have previously discussed. However, as this only effects tensor contractions and not the much more costly eigenvalue search in the DMRG algorithm, the impact on the runtime remains manageable.
Nevertheless, it remains to be seen how DMRG will navigate these intricate constraints and what implications they may have on the solution's accuracy.

\subsubsection{An MPO representation for balanced laminates}

As outlined in section \ref{sec:constraints}, the constraint for balanced laminates, as well as generalized constraints where two states, $s$ and $t$, appear an equal number of times in a stacking sequence, can be expressed using a penalty operator:
\begin{equation}
    \hat{H}_{\text{balanced}} = \gamma \qty(\sum_{n=1}^N \hat{H}^{[n]})^2 ,
\end{equation}
where the local operators $\hat{H}^{[n]}$ are defined for each subsystem $n$ as:
\begin{equation}
    \hat{H}^{[n]} = \dyad{s}_n - \dyad{t}_n = \sum_{s_n=1}^d \qty(\delta_{s_n,s} - \delta_{s_n,t}) \dyad{s_n}_n.
\end{equation}
Given that the penalty operator bears similarity to the Hamiltonian form for the lamination parameters, as seen in eq. (\ref{eq:ham_xl}), the same methodology discussed in section \ref{sec:mpoloss} can be applied. For this, we set the eigenvalues $h^{[n]}_{s_n}$ of the system to:
\begin{equation}
    h^{[n]}_{s_n} = \delta_{s_n,s} - \delta_{s_n,t}, 
\end{equation}
yielding a representation of the penalty operator $\hat{H}_\mathrm{penalty}^{\text{(balanced)}}$ in MPO form. The factor $\gamma$ can be included by multiplying it to any one of the local tensors.

\subsubsection{An MPO representation for the 10\%-rule}

As detailed in section \ref{sec:constraints}, constraints such as the $10\%$-rule require a minimum of $N_s$ plies to be in a specific state $s$. For this constraint, we implement a linear penalty function:
\begin{equation}
    H_{\text{$10\%$-rule}}(\vec{s}) =  \begin{cases}
    0, & \text{if } \sum_{n=1}^N \delta_{s_n,s} \geq N_s,\\
    \gamma \qty(N_s - \sum_{n=1}^N \delta_{s_n,s}), & \text{else,}
\end{cases},
\end{equation}
where the parameter $\gamma>0$ controls the magnitude of the added penalty. While an expansion of the according operator into single-site operators is more intricate, as discussed in section \ref{sec:constraints}, a straightforward representation as an MPO can be found via the decomposition:
\begin{equation}
    H_{\text{$10\%$-rule}}(\vec{s}) = \begin{bmatrix}
    1 & 0 & \cdots & 0  
\end{bmatrix} M(s_1) \cdots M(s_N) \begin{bmatrix}
    \gamma N_s \\ -\gamma \\ \vdots \\ -\gamma 
\end{bmatrix}, \label{eq:penalty10percentdecomp}
\end{equation}
where each $M(s_n)$ is a square matrix of dimension $N_s+1$. For $s_n = s$, $M(s)$ is defined as:
\begin{equation}
    M(s) = \left[\begin{array}{c@{\hspace{0.65cm}}cccc}
     1 & 1 & 0 & \cdots & 0 \\
     0 & 0 & 1 & \cdots & 0 \\
     \vdots & \vdots & \ddots & \ddots & \vdots\\
     0 & 0 & \cdots &  0 & 1 \\[0.1cm]
     0 & 0 & \cdots &  0 & 0
\end{array}\right],
\end{equation}
and for $s_n \neq s$, it is equivalent to the identity matrix:
\begin{equation}
    M(s_n) = \textbf{I}_{N_s+1} \qq{for} s_n \neq n.
\end{equation}
Consequently, multiplying $M(s_n)$ with a vector $\vec{a}\in \C^{N_s + 1}$ leaves $\vec{a}$ unchanged if $s_n \neq s$. However, for $s_n = s$, it shifts all elements below the first row up by one position, combines the first two elements in the first row, and appends a zero at the end:
\begin{equation}
    M(s) \begin{bmatrix}
    a_1 \\ a_2 \\ \vdots \\ a_{N_s+1}
\end{bmatrix} = \begin{bmatrix}
    a_1 + a_2 \\ a_3 \\ \vdots \\ a_{N_s+1} \\ 0
\end{bmatrix}.
\end{equation}
Thus, applying this matrix to the column vector in eq. (\ref{eq:penalty10percentdecomp}) cumulatively adds $-\gamma$ for each occurrence of $s_n = s$, effectively subtracting from the penalty. If the constraint is satisfied for $\vec{s}$, the penalty becomes zero after $N_s$ such additions. Meanwhile, the remaining elements of the vector are now also 0, such that multiplying any further matrices leave the vector unchanged. Conversely, if the constraint is not met, the penalty remains non-zero, reflecting the penalty function $H_{\text{$10\%$-rule}}(\vec{s})$. This justifies the decomposition in eq. (\ref{eq:penalty10percentdecomp}), which can be used to construct the MPO similarly to previously discussed cases.

Similar to the contiguity constraint, these matrices are significantly larger than those of the other encountered MPOs. As discussed earlier, this should have a comparatively minor impact on the runtime, while the influence on the accuracy remains to be explored.

\subsection{The connection between the hardware-efficient approach and DMRG}

In this section, we highlight the similarities between the hardware-efficient approach and the DMRG algorithm, and contrasting these with the properties of QAOA. From an implementation perspective, both QAOA and the hardware-efficient approach involve encoding states in the Hilbert space on qubits, performing state-vector simulations with a designated number of repetitions $N_r$ of a parameterized quantum circuit, and extracting the Hamiltonian's expectation value. A classical optimizer is then employed to search for parameters that minimize this value. However, the specific state preparation circuits and optimization procedures lead to different landscapes for the Hamiltonian expectation value. Specifically, QAOA utilizes operators $U_C$ and $U_M$ that act simultaneously on all qubits with a single parameter per operator, allowing for the simultaneous optimization of a relatively small number of parameters. In contrast, the hardware-efficient approach optimizes only the parameters affecting neighboring qubits, much like how DMRG optimizes subsystems for neighboring plies.

Furthermore, the connection between the hardware-efficient approach and DMRG extends deeper \cite{Rudolph2022a, Rudolph2022b,Ran2020,Malz2024}. Viewing the gates in a quantum circuit as matrices naturally leads to a tensor network representation of the circuit. The initial qubit states $\ket{0}$ can be included in vector form to form a tensor network representing the states prepared by the circuit. In the hardware-efficient circuit, where two-qubit gates act only on neighboring qubits, performing singular value decompositions (SVD) on these gates and contracting all tensors associated with individual qubits effectively yields an MPS representation of the prepared states. This results in open indices for individual qubits, but by grouping the qubits associated with one ply, we can also achieve an MPS representation similar to those used in DMRG. This conceptualization allows us to view the hardware-efficient approach as a qubit implementation of a specifically parameterized MPS.

The bond dimension of this MPS can be inferred from the SVD of the CNOT gate. Typically, the CNOT is represented as a matrix with non-zero elements $C_{q_1 q_2, q_1' q_2'} := \mel{q_1 q_2}{\mathrm{CNOT}}{q_1' q_2'}$:
$$ C_{00,00} = C_{01,01} = C_{10,11} = C_{11,10} = 1,$$
which are applied to the current two-qubit state and return the state after the application of the gate. However, performing an SVD on this matrix would split the matrix along the time direction, and not separate the qubits. We therefore reorder the indices $\tilde{C}_{q_1 q_1',q_2 q_2'} = C_{q_1 q_2,q_1' q_2'} $ to obtain non-zero elements:
$$ \tilde{C}_{00,00} = \tilde{C}_{00,11} = \tilde{C}_{11,01} = \tilde{C}_{11,10} = 1. $$
Notably, the CNOT gate does not alter the control qubit, leading to a rank-2 matrix representation. Consequently, the SVD yields only 2 non-zero singular values, implying that each CNOT gate between two qubits doubles the bond dimension of the corresponding MPS. As each repetition in the hardware-efficient circuit introduces an additional CNOT gate between two qubits, the bond dimension of the MPS grows exponentially with $2^{N_r}$. With this result, we can compare this hardware-efficient circuit directly to an MPS of a specific bond-dimension that is utilized in DMRG. Conversely, QAOA uses numerous CNOT gates per repetition on non-neighbouring qubits, stemming from the many terms in the Hamiltonian. Consequently, the tensor-network representation of the circuit has a significantly more complicated form than a simple MPS. We can therefore expect the harware-efficient approach and DMRG to have comparable behaviours, while QAOA will likely perform very differently.

However, there are key differences between the hardware-efficient approach and DMRG. The hardware-efficient approach optimizes neighboring qubits, whereas DMRG directly operates on the ply subsystems $\calH_n$. In principle, the optimization in the hardware-efficient approach could also be performed by simultaneously optimizing the 4 qubits corresponding to 2 plies, which introduces the challenge of handling more parameters concurrently. Moreover, while the hardware-efficient approach utilizes a parameterized MPS with a classical optimizer, DMRG conducts optimization through eigendecomposition of an environment tensor followed by SVD to optimize local tensors. This method also impacts the control of bond dimensions, since this procedure naturally increases the bond-dimension and offers tools for its controlled reduction, unlike in the hardware-efficient approach where the bond dimension remains fixed and a method to reduce the number of repetitions $N_r$ is not readily available. This analysis underscores how the exploration of tensor-network methods like DMRG can provide valuable insights into specific quantum algorithms such as our hardware-efficient approach.

%% file: V_demo.tex
\section{Numerical Demonstration} \label{sec:demo}

We performed numerical simulations of the algorithms outlined in \cref{sec:algo} to demonstrate our quantum formulation of stacking sequence retrieval as presented in \cref{sec:quantum}. Specifically, we conducted state-vector simulations for the two quantum algorithms discussed in \cref{sec:vqa}: QAOA and the previously described hardware-efficient approach. As DMRG is a classical algorithm, it did not require simulation on quantum computing hardware and was implemented directly.

Given the limitations of simulating quantum computing on classical hardware, particularly with a large number of qubits, we limited our simulations to a symmetric laminate comprising a small number of $N=6$ plies in one half of the stack, corresponding to a total of 12 simulated qubits. However, the MPO and MPS representations used in DMRG allow for a more efficient management of the state space compared to the direct simulation of subsystems, enabling DMRG to efficiently handle a much larger number of subsystems. Consequently, we also conducted DMRG simulations for a substantially larger configuration of $N=200$ plies.

All simulations were performed both with and without the inclusion of a disorientation constraint of $45^\circ$.

\subsection{Setup}


The numerical calculations were carried out on a laptop equipped with an Intel Core i7-1185G7 processor (4 cores, 3.00 GHz), 16GB RAM, running Windows 10 Professional. The complete code for these experiments is available at \cite{Wulff2024code}. The algorithm was executed on the CPU without using parallelization techniques like multi-threading. All files generated from our experiments are available in HDF5 format at \cite{Wulff2024data}.

The quantum simulations were performed in \textit{Python} in version 3.11 using the \textit{Qiskit} library in version 1.0.2 \cite{qiskit2024}. This library facilitated the construction and simplification of the loss function and penalty Hamiltonian in expanded Pauli operator form, as detailed in \cref{sec:qaoa}. Additionally, the state preparation circuits for QAOA and the hardware-efficient approach were generated as quantum circuits, which \textit{Qiskit} converts into an instance of a state vector class. This class provides a function for calculating the expectation value of operators such as the Hamiltonian. Optimization was conducted using the \textit{Scipy} library in version 1.12.0 \cite{scipy}, employing the BFGS method as the optimizer.

The numerical implementation and trials were executed in the programming language \textit{Julia}, version 1.8.4. The MPS and MPO were implemented using the \textit{ITensors.jl} library, version 0.3.22 \cite{itensor,itensor-r0.3}, which also includes an implementation of the DMRG algorithm. As this implementation supports only alternating sweeping directions, we modified the algorithm to allow for custom sweeping sequences. Additional functionality for recording data during the optimization was also integrated. Aside from the support for custom sweeping directions, the optimization process remains consistent with the original implementation provided by \textit{ITensors}.


For all numerical trials, we selected the four commonly used angles $0^\circ,+45^\circ,90^\circ,-45^\circ$, as discussed in \cref{sec:algo}. In the quantum simulations, these angles were encoded using binary encoding as described in section \cref{sec:vqa}. We applied all three algorithms to a symmetric laminate with $N=6$ plies in one half of the stack and optionally included a disorientation constraint. For this constraint, we applied a penalty of $\gamma=0.25$ per constraint violation (sections \cref{sec:constrnn} and \cref{sec:mponn}). Additionally, we performed DMRG on a larger scale with $N=200$ plies in half the stack, assigning a penalty of $\gamma = 0.005$. All trials were conducted with and without the disorientation constraint.


To select a well-justified set of target lamination parameters, we considered two key factors. Firstly, we chose target parameters that correspond to exact solutions of feasible stacking sequences. This approach ensures that the loss function serves as an effective quality measure for any obtained solution, since an optimal solution with a minimal loss function value of $0$ always exists. Suitable target parameters can be obtained from generated stacking sequences that comply with the constraints, such that these sequences also act as known optimal solutions for the target parameters.


Secondly, it is crucial that the target lamination parameters represent a diverse cross-section of potential lamination parameters. For $N=6$ layers, we achieved this by considering symmetry transformations of the constraint. As we discuss further in \Cref{sec:app_sym}, we anticipate that the algorithms will exhibit similar behavior in cases that transform into each other under transformations that map valid states to valid states and constraint-violating states to constraint-violating states. For the disorientation constraint, these symmetry transformations are generated by rotations that shift all states by the same number and one reflection that swaps 0 and 2, or equivalently 1 and 3. To explore a broad range of potential behaviors, we generated 15 stacking sequences that do not transform into each other under these symmetry transformations, resulting in a set of 15 distinct target lamination parameters for our trials with $N=6$ plies.


For $N=200$, the total number of valid stacking sequences is vastly greater than for $N=6$ plies, necessitating a different approach. Common methods for generating random stacking sequences often result in density variations within the corresponding point cloud in the 8-dimensional lamination parameter space, typically due to the specifics of the sequence generation method. To mitigate potential biases from our random generation method, we compiled a large dataset of 500,000 stacking sequences that comply with the disorientation constraint, subsequently converting these into an equivalent number of points in lamination parameter space. We then applied a kernel density estimator to evaluate the point density at each location \cite{Rosenblatt1956,Parzen1962,Weglarczyk2018}. Finally, we used the inverses of these densities as weights to randomly select points in a manner that more closely approximates a uniform distribution across the lamination parameter space. Employing this method, we sampled a set of 50 target lamination parameters for all our trials with $N=200$ plies.


For the QAOA trials, we conducted simulations for $N_r = 1, 2, 3, 4, 6, 8$ repetitions of the problem and mixer unitaries in the state preparation circuit. For each number of repetitions $N_r$, we generated 6 random sets of initial parameters, $\beta$ and $\gamma$, which were used across all 15 target lamination parameters.

In the hardware-efficient approach, we performed optimizations for $N_r = 0, 1, 2, 3, 4, 5, 6$ repetitions in the state preparation circuit. For each repetition count, we generated 5 random sets of initial parameters. Additionally, for each combination of repetition number, initial parameters, and target, we conducted the optimization twice: once sweeping outward (from $n=1$ to $n=N$) and once sweeping inward (from $n=N$ to $n=1$). Multiple sweeps were performed consecutively until convergence was achieved.


For DMRG, we conducted optimizations for maximum bond dimensions of 2, 4, 8, 16, and 32. We performed optimizations using alternating sweeps, as is customary in DMRG, as well as only inward and only outward sweeps. In the final stages of optimization, we applied a technique to enforce a basis state corresponding to a single stacking sequence, which involved progressively halving the maximum bond dimension during the last few sweeps, ultimately achieving a bond dimension of 1 in the penultimate sweep. In the final sweep, we applied a truncation criterion that discarded singular values below 0.5 during the dimension reduction, effectively enforcing the representation of a basis state.

For $N=6$, we based the number of sweeps for DMRG on the quick convergence observed in the hardware-efficient approach, as discussed in the following section \cref{sec:results}. Ensuring sufficient sweeps to both grow and reduce the bond dimension as described above, we executed DMRG with a total of 10 sweeps. For $N=200$, to better observe the optimization progression over a more extended series of sweeps, we opted for 69 sweeps, yielding a total of 70 data points per trial when including the initial state.

For each target lamination parameter, we generated 5 initial MPS with a bond dimension of 2. These initial MPS were consistently used across all trials, irrespective of bond dimension and sweep sequence. Given that the initial bond dimension was lower than most of the chosen maximum bond dimensions, the bond dimension of the MPS increased quickly during the initial sweeps of optimization until it reached the predetermined maximum. Therefore, the majority of the sweeps were conducted at this maximum bond dimension. 

\subsection{Results} \label{sec:results}

\begin{figure}[t!]
    \centering
    \includegraphics[scale=1.0]{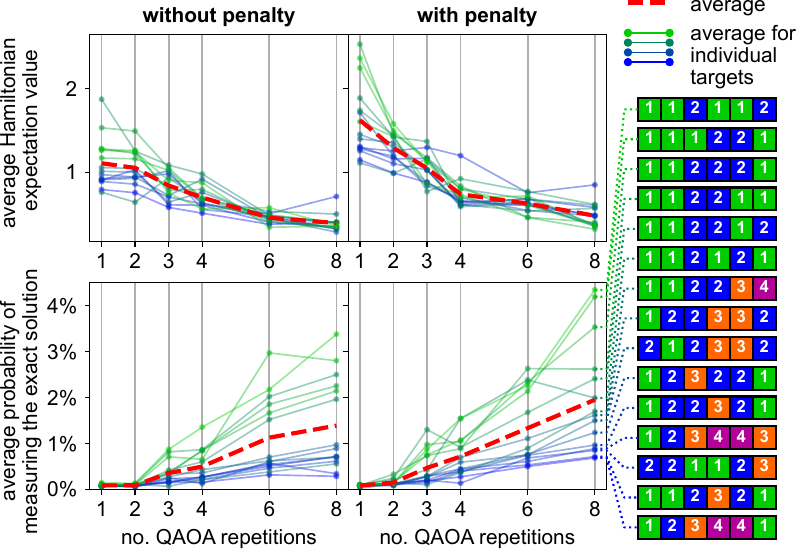}
    \caption{Results from the simulations of QAOA. The top row shows the average Hamiltonian expectation value while the bottom row shows the average probability of obtaining the exact solution when measuring the final state. Shown are the average of the 6 trails for each individual set of target parameters (blue and green solid lines) and the average over all 15 sets of target parameters (dashed red lines). On the right, the stacking sequences corresponding to the exact solutions are shown.}
    \label{fig:qaoa}
\end{figure}

\begin{figure}[t!]
    \centering
    \includegraphics[scale=1.0]{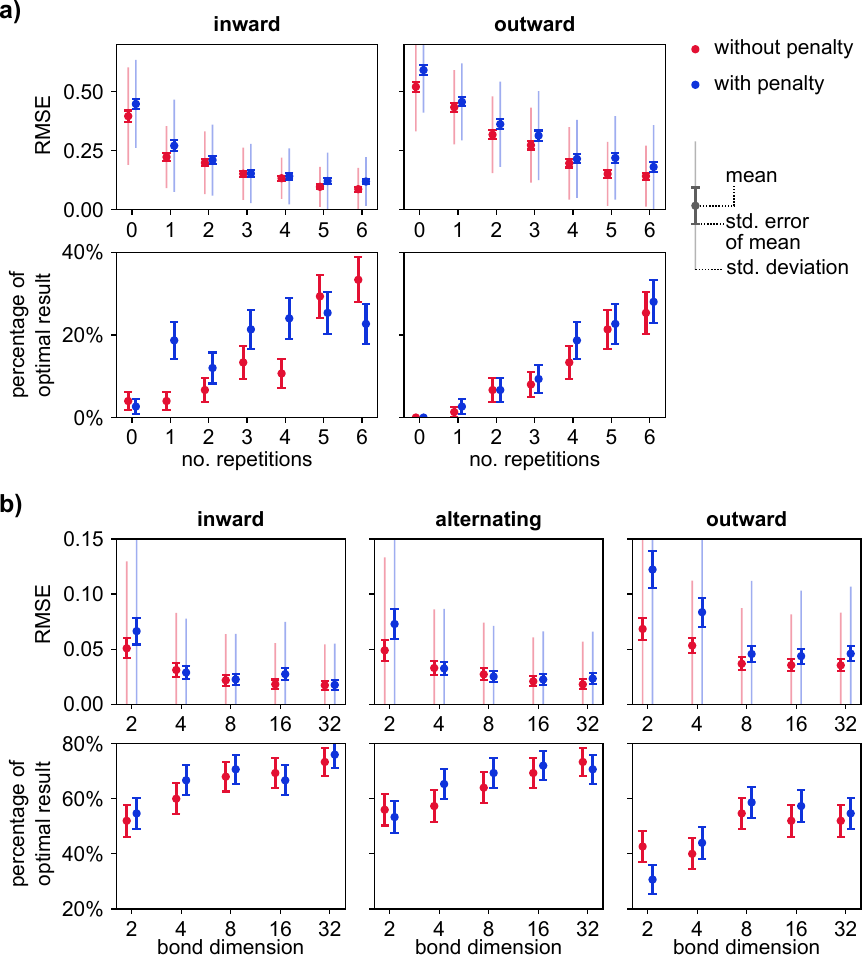}
    \caption{Results for the hardware-efficient approach \textbf{(a)} and DMRG \textbf{(b)}. In both cases, the top row shows the average RMSE for the final results, and the bottom row shows the ratio of resulting stacks that coincided with the exact solution. The averages are taking over the results from all 15 sets of target lamination parameters with 5 trials performed for each. }
    \label{fig:hwe_dmrg_6}
\end{figure}

\begin{figure}[t!]
    \centering
    \includegraphics[scale=1.0]{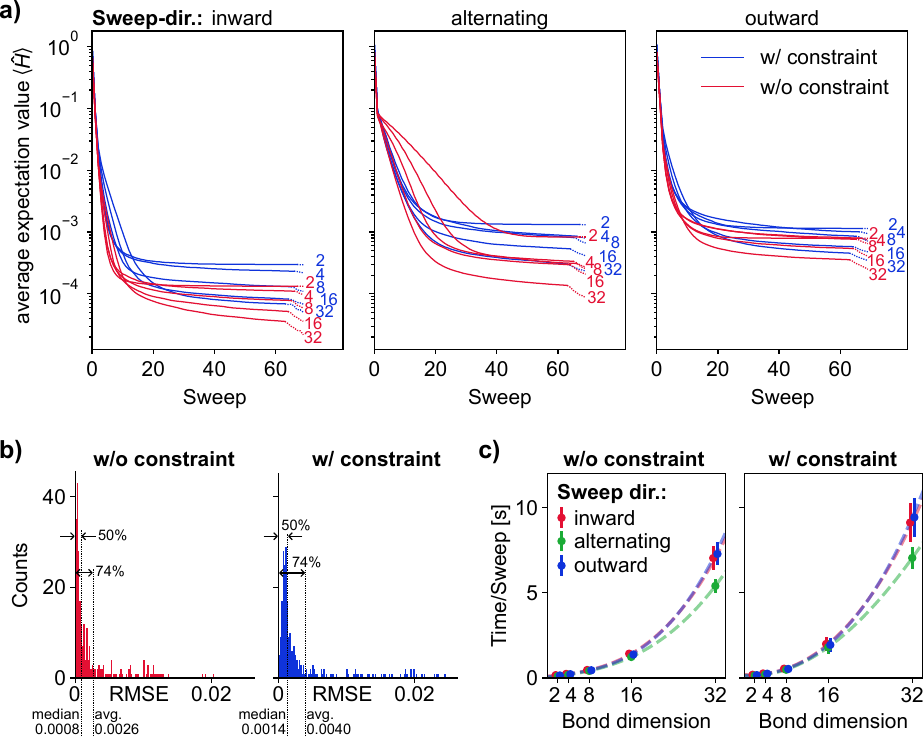}
    \caption{Results from DMRG with $N=200$ plies. \textbf{a)} Traces of expectation values of the loss function Hamiltonian, averaged over all samples and measurements after each sweep. Each line corresponds to a different bond dimension, denoted at the end of each trace. Red lines represent trials excluding the disorientation constraint and blue lines correspond to trials including the constraint. Solid lines represent the optimization phase with a constant maximum bond dimension. The successive reduction of the bond dimension at the final phase of the optimization is indicated by dotted lines that can be found at the right-hand end of each trace. \textbf{b)} Histogram of the RMSE of the solutions for maximum bond-dimension 32 and inward direction. Counted were the points from all target lamination parameters and initial MPS. The bin-size is $0.002$. The ranges of RMSE for the bottom half of the points, and up to the average RMSE are shown in both plots.  Left: Without the disorientation constraint. Right: With the disorientation constraint. \textbf{c)} Average sweep duration as a function of bond dimension. Left and right plots respectively correspond to scenarios without and with the disorientation constraint. Each sweep direction is represented by distinct points in different colors. A third-order polynomial fit is included as dashed lines, to easier distinguish trends for different configurations.}
    \label{fig:dmrg200}
\end{figure}


The general concept of QAOA is to exploit quantum interference to suppress the amplitude of states associated with high losses and enhance the probability of measuring states with lower losses. Accordingly, our trials with QAOA did not yield a specific basis state corresponding to a particular stacking sequence, but resulted in a superposition of states encompassing a range of losses. For each set of resulting QAOA parameters $\vec{\beta}$ and $\vec{\gamma}$, we calculated the loss function expectation value of the state and assessed the probability of obtaining the exact solution from the corresponding component of the state. The results are presented in \Cref{fig:qaoa}, where we display the Hamiltonian expectation values and the probabilities of obtaining the exact state separately, comparing results with and without the inclusion of the penalty operator. In each plot, we display the overall average for each number of repetitions $N_r$ as well as individual traces for each of the 15 targets, averaging only over the 6 runs with different initial parameters. The figure also includes the stacking sequences corresponding to the target lamination parameters. Additionally, we calculated the expectation value of the penalty operator for the disorientation constraint, setting the penalty factor to $\gamma = 1$, thus the recorded expectation values represent the average number of constraint violations in the stack. For $N_r = 1$ repetition in the QAOA circuit, we found an average expectation value of $0.309$ constraint violations without the penalty operator, and $0.261$ with it. As the number of repetitions $N_r$ increased, these values decreased, with $0.184$ and $0.085$ recorded for $N_r = 8$ repetitions, respectively, without and with the penalty operator.


For the hardware-efficient approach, the behavior observed was markedly different from that of QAOA. Specifically, the hardware-efficient approach effectively produced a basis state after the initial sweep, with subsequent sweeps primarily filtering out remnants of other states. The results are depicted in \Cref{fig:hwe_dmrg_6}a, where we show the average RMSE to the target parameters for the final stacking sequences and the percentage that coincided with the optimal state. Additionally, we present results for sweeping inward and sweeping outward in separate plots. No clear trend emerged for different target lamination parameters; thus, we only show the average of all results for a specific number of repetitions $N_r$, with and without the constraint included in the same plot. We also tracked the number of resulting stacking sequences that violated the disorientation constraint. Without the penalty, inward sweeps at $N_r = 0$ repetitions resulted in $65\%$ constraint-violating stacks, while subsequent repetitions yielded between $24\%$ and $39\%$ constraint-violating stacks without a discernible trend with increasing repetitions. Similarly, outward sweeps resulted in $40\%$ constraint violations at $N_r = 0$ and values between $24\%$ and $39\%$ for other numbers of repetitions, also without showing a clear correlation to the inward sweep values. Including the penalty, we observed $29\%$ and $27\%$ constraint violations for inward and outward sweeps respectively at $N_r = 0$. However, with an increase in repetitions $N_r$, the percentage of constraint-violating stacks rapidly decreased, dropping below $1.5\%$ at $N_r = 2$ and $N_r = 3$ for inward and outward sweeps respectively.


The reduction of the maximum bond dimension for the final sweeps in our implementation of DMRG is designed to enforce a basis state as a result. For $N=6$, we display the results for DMRG in \Cref{fig:hwe_dmrg_6}b in a similar fashion as for the hardware efficient approach. The main difference is, that we now have three different sweeping configurations: inward, outward and alternating sweeps. As before, we also counted the states containing constraint violations. Without the penalty, inward and alternating sweeps yielded $19 \%$ and $17 \%$ constraint violating stacks respectively for bond-dimension 2, which reduced to $12 \%$ in both cases for bond-dimension 32. Outward sweeps showed higher values, with $28\%$ for bond-dimension 2 and $24 \%$ for bond-dimension 32. When including the penalty, we only observed a single constraint-violating stack for alternating sweeps and a single one for outward sweeps, both at bond-dimension 2, which equates to $1.3 \%$.

For DMRG trials with $N=200$ plies, we offer a more comprehensive overview of performance. Firstly, we confirmed that all resulting stacking sequences were free of constraint violations when the disorientation constraint was included. Without the penalty, every combination of maximum bond dimension and sweeping direction yielded at least $96\%$ constraint-violating stacks, indicating that nearly all resulting stacks violated the disorientation constraint.

Additionally, we documented the expectation value of the loss function Hamiltonian after each sweep in all trials. \Cref{fig:dmrg200}a shows the average across all target lamination parameters as a function of the number of sweeps. Each trace corresponds to a different configuration regarding bond dimension, sweeping direction, and whether or not the disorientation constraint was included. For a more detailed discussion on the distribution of the obtained solutions, \Cref{fig:dmrg200}b provides a histogram of the Root Mean Square Error (RMSE) across all samples for the most effective configuration (bond dimension 32 and sweeping left), both with and without the disorientation constraint. The bin size of the histograms is set to 0.002. A comprehensive overview of the results from all individual trials for DMRG with $N=200$ plies is available in \Cref{fig:dmrg_samples} in Appendix \ref{sec:app_allsamples}.

We also monitored the time duration of each sweep. \Cref{fig:dmrg200}c illustrates the average sweep duration as a function of the bond dimension, with separate traces to distinguish between different sweeping directions and comparisons of cases with and without the disorientation constraint. For calculating the averages and standard deviations, the first and last 10 sweeps were excluded. To minimize the impact of background processes on the operating system on the duration of isolated sweeps, the lowest and highest 10\% of times for all considered sweeps were also excluded. In the DMRG algorithm, the most computationally intensive step involves finding the eigenvector with the lowest eigenvalue of a matrix, which is a critical component of the local optimizations of the site tensors. The computational complexity for solving the eigenvalue problem generally scales as $O(b^3)$, where $b$ is the bond dimension. This scaling behavior is consistent with that of the Lanczos algorithm used for the eigenvalue computations, assuming the number of non-zero matrix elements scales proportionally to the overall size of the matrix. However, due to the limited number of data points for different bond dimensions, definitive assertions about the actual observed scaling behavior are not possible. Nonetheless, for illustrative purposes, a third-order polynomial fit has been included in the figure to facilitate easier comparison of the different configurations.

\subsection{Discussion}

We will first discuss and compare the chosen methods for $N=6$ plies, then provide a more detailed analysis of DMRG for $N=200$ plies.

With $N=6$ plies, we observed for all three algorithms that increasing the number of repetitions $N_r$ in the state-preparation circuit or raising the bond dimension in DMRG generally leads to better outcomes in terms of loss or RMSE, and the likelihood of obtaining the exact state. Comparing Figures \ref{fig:qaoa} and \ref{fig:hwe_dmrg_6}, the hardware-efficient approach and DMRG appear to significantly outperform QAOA. Furthermore, QAOA is considerably more expensive to execute due to its substantially larger circuit depth. However, as the hardware-efficient approach and DMRG generally end up in a basis state, which often has insufficient accuracy according to the large standard deviations in \Cref{fig:hwe_dmrg_6}, multiple trials with different initial parameters may be necessary to obtain a good solution \cite{McClean2018,Zhou2020,Grant2019,Verdon2019}. Conversely, the final superposition of states in QAOA generally includes good solutions, which can be found with a sufficient number of measurements of the final superposition, making it less expensive than multiple trials of the entire optimization. Nevertheless, QAOA faces well-known issues like barren plateaus that may still require multiple trials and optimized strategies \cite{Zhu2022,Khairy2020,Bravyi2021,Patel2024}. Additionally, we chose basic VQA examples for our demonstrations, and as more advanced methods are developed, further research will determine which strategies yield the best results when scaling up the number of plies to industry-relevant levels.

It is also important to note that we used exact state-vector simulation for the quantum algorithms. However, unlike DMRG, which is a classical algorithm, executing quantum algorithms on quantum hardware introduces various errors such as state-preparation and measurement (SPAM) errors and various noise channels during the execution of a quantum circuit \cite{Preskill2018}. Investigating strategies to manage these noise factors is a major research area, and future studies must address how these errors and mitigation strategies impact the design and effectiveness of quantum algorithms for the SSR problem. 

We continue with a discussion of the results for $N=200$ plies. For $N=6$ plies, the solution space consists of $d^N = 4096$ configurations, making it feasible to find the optimal solution. However, for larger numbers of plies, the configuration space expands exponentially, shifting the focus from identifying the optimal solution to finding a highly accurate solution. While current quantum hardware and quantum simulations are generally limited by the system size over which algorithms can be effectively executed. The scalable DMRG algorithm however can provide insights that are applicable to some VQA like the chosen hardware-efficient approach.

As shown in \Cref{fig:dmrg200}a, the high number of sweeps allow for studying the progression of the Hamiltonian expectation value during the optimization. In all configurations, there is a rapid decrease in the expectation value during the initial sweeps, followed by a significant slowdown in the optimization rate. 
The reduction in bond dimension, indicated by the dotted lines at the end of each trace in \Cref{fig:dmrg200}a, leads to a pronounced reduction in the expectation value of the loss function. This indicates that just before this reduction, the MPS represents a superposition of basis states that includes less optimal states, which are then filtered out during the bond dimension reduction.

While the averaged traces provide an overview, they do not reveal the actual distribution of all resulting lamination parameters for each specific configuration. The histograms in \Cref{fig:dmrg200}b and \Cref{fig:dmrg_samples} demonstrate how the majority of the trial significantly outperform the worst. Furthermore, \Cref{fig:dmrg_samples} shows that the performance does not necessarily correlate between different configurations in initial MPS, bond-dimension and sweeping direction. Consequently, warm-starting and multiple trials with various configurations can potentially enhance the accuracy of the result significantly.

Comparing the loss function expectation values of the results together with the duration of a sweep in \Cref{fig:dmrg200}c for distinct bond dimensions shows a clear trade-off between accuracy and runtime. The main contributor to the runtime is the eigen-decomposition, which has a cubic dependence on the bond dimension. This increase makes DMRG impractically demanding for high bond dimensions. In contrast, we only need a logarithmic increase in the circuit depth $N_r$ in the hardware-efficient circuit to grow the effective bond dimension. However, this does not yet consider the optimization procedure, and the large number of parameters to be optimized might counteract this gain.

Considering the practical aspects of our research approach, it is crucial to highlight that our numerical trials were conducted using the DMRG method, based on the implementation provided by the \textit{ITensors.jl} library, on a CPU without optimizing for performance. Implementing parallelization techniques, GPU support, sparse matrix optimizations, or problem-specific methods could lead to substantial performance enhancements. Additionally, adjusting other DMRG settings, such as hyperparameters in the eigenvalue solver, might further impact performance. Furthermore, to take advantage of the varying performance of different operationial configurations, techniques like warm-starting can potentially reduce the requirements for finding a good solution in the majority of cases \cite{Truger2023}.

In conclusion, our study lays the groundwork for further investigation into the potential of DMRG as a viable alternative for stacking sequence retrieval. To achieve this, it is essential to incorporate the specific manufacturing constraints discussed earlier, such as the contiguity constraint and the $10\%$-rule, which are crucial for real-world applications \cite{Niu1988,Bailie1997,compmathandbook2002}. Alongside these additions, implementing performance enhancements such as parallelization, GPU support, and optimization of algorithm settings is critical for this method to be competitive with existing stacking sequence retrieval techniques. 
Nevertheless, the results from our study offer encouraging signs that tensor networks and quantum algorithms could play a significant role in advancing quantum and quantum-inspired methods for laminate design.

\FloatBarrier

%% file: VI_conclusion.tex
\section{Conclusions and Outlook} \label{sec:conclusion}

In this work, we have developed novel approaches based on quantum computing and DMRG to tackle the stacking sequence retrieval problem, a notable challenge in the field of laminated composite design.

We initiated our study by developing a quantum representation of the SSR problem, adapting it for quantum computing methods. Initially, the problem was formalized as an integer optimization problem, where ply-angle sequences were represented as a list of integers, and the loss function was defined to quantify the mean square error between the lamination parameters of a given configuration and the target parameters. These stacking sequences were subsequently embedded in a Hilbert space, with each basis state corresponding to a distinct stacking sequence, alongside a Hamiltonian that encapsulated the loss function as a Hermitian operator. Various manufacturing constraints and their implementation as penalty terms in the loss function were also examined. The translation of these penalty terms into Hamiltonian operators revealed diverse forms of subsystem interactions across individual plies.

We then proceeded to validate our approach through numerical simulations. For this purpose, we selected two variational quantum algorithms, QAOA and a hardware-efficient approach for which we performed state-vector simulations. The latter shares similarities with the classical DMRG algorithm which we selected as a third algorithm for out study. We explored relevant concepts regarding the implementation of these algorithms and assessed their impact on algorithm performance. Simulations were conducted for a modest number of $N=6$ plies, which exhibited distinct behaviors between QAOA and notable similarities between the hardware-efficient approach and DMRG. The efficient implementation of DMRG further facilitated simulations for a larger laminate with $N=200$ plies. Each algorithm demonstrated varying degrees of success in identifying good solutions, affirming that our quantum formulation of the SSR problem is a viable approach.

While our study on a single panel was conducted with the future goal of addressing more complex scenarios involving multiple panels, the results from employing the well-known DMRG algorithm in this new scenario are promising even for a single panel. However, to establish its competitiveness, several key aspects need to be addressed in future work. A critical aspect is the integration of additional manufacturing constraints, which are vital for real-world applications, and evaluating whether the DMRG algorithm can efficiently manage these complexities. Moreover, implementing performance optimizations will be crucial in assessing whether this approach can rival existing classical methods in efficacy.

A thorough comparison with existing state-of-the-art algorithms will determine whether DMRG is a serious contender. Such comparisons might also reveal advantages of DMRG's heuristic nature that could benefit specialized SSR strategies. For example, employing MPS as in DMRG allows for efficiently representing a superposition of stacking sequences, which could enable novel specialized strategies for multi-panel designs. An MPS consisting of approximate solutions can be obtained, for instance, by stopping the DMRG algorithm early before convergence. Future approaches might also explore more sophisticated tensor network configurations for multi-panel laminates, making tensor network methods a contender to quantum algorithms in these complex cases. Conversely, a collaborative exploration of tensor networks and quantum algorithms may unveil new insights, propelling advancements that leverage the strengths of both methodologies.

The quantum algorithms selected for our simulations, though basic, are valuable for validating our approach and represent just two examples within a broad spectrum of existing and potential quantum computing methods, with new specialized strategies being developed rapidly. Therefore, our simulations offer only a limited view of the future potential of quantum methods for SSR. Nevertheless, a key motivation of this work is to establish it as a promising candidate for industry-relevant applications of quantum computation. By presenting our findings to the broader community, we aim to stimulate the development of quantum algorithms tailored to industry-relevant problems, thereby fostering real-world impact.

Our exploration into the application of quantum computing for stacking sequence retrieval, particularly the process of selecting optimal sequences from a set of discrete choices, may have implications beyond laminated composite design. For example, the definition of the lamination parameters in \cref{sec:ssrinteger} resembles the weighted-sum model in multi-criteria decision making, which is employed in a variety of applications \cite{Fishburn1967,Chruchman1954,Figueira2005}. 
This suggests that the methodologies and insights gained from our study can potentially inform approaches to quantum computing in other domains, thereby contributing to the progress in the practical application of quantum computational methods.

In conclusion, this research establishes a foundation for a further exploration into the utility of quantum algorithms for stacking sequence retrieval and laminate design, by developing a quantum representation of the problem. Moreover, we introduced tensor networks as a potential quantum-inspired approach. Importantly, our numerical results substantiate the efficacy of our methods, and provide valuable insights into benefits and drawbacks of quantum and quantum-inspired methods for stacking sequence retrieval. Continued research is crucial to fully understand the role these advanced computational techniques might play in the future of laminate design.

%% file: formalities.tex
\section*{Author contributions}

\textbf{Arne Wulff:} Conceptualization, Methodology, Formal Analysis, Investigation, Software, Data curation, Writing - original draft, Writing - review \& editing, Visualization, Validation, 
\textbf{Boyang Chen:} Conceptualization, Writing - review \& editing, Supervision, Funding acquisition,
\textbf{Matthew Steinberg:} Methodology, Writing - review \& editing,
\textbf{Yinglu Tang:} Writing - review \& editing, Supervision, Funding acquisition,
\textbf{Matthias M\"oller:} Writing - review \& editing, Supervision, Funding acquisition,
\textbf{Sebastian Feld:} Writing - review \& editing, Supervision, Funding acquisition

\section*{Code and data availability}

The code for performing the experiments in \cref{sec:demo} can be found in a \href{https://github.com/ArneWulff/stacking-sequence-retrieval-with-dmrg}{GitHub repository} \cite{Wulff2024code}. The generated data is available can be found in on \href{https://doi.org/10.4121/ae276609-55b0-4af1-88c0-1102b1b58990}{4TU.ResearchData} \cite{Wulff2024data}.

\section*{Acknowledgements}

We thank the remaining members of the QAIMS lab, Swapan Madabhushi Venkata, Koen Mesman and Philip W\"urzner, for valuable discussions regarding this work. We thank the Faculty for Aerospace Engineering at the Delft University of Technology for their funding support of the QAIMS lab.

\section*{Declaration of Interests}

The authors declare that they have no known competing financial interests or personal relationships that could have appeared to influence the work reported in this paper.

\section*{Declaration of generative AI in scientific writing}

During the preparation of this work the authors used OpenAI's ChatGPT exclusively for language suggestions and corrections. After using this service, the authors reviewed and edited the content as needed and take full responsibility for the content of the publication.

%% file: appendix.tex
\appendix
\section{Appendix}

\subsection{Terms in the penalty for the 10\%-rule} \label{sec:app_penalty10perc}

In this section, we have a closer look into penalties to ensure the $10\%$-rule, as discussed in section \ref{sec:constraints}. There, in \cref{eq:penalty10perc}, we defined a possible penalty function to depend on the number of missing plies to satisfy the constraint:
\begin{equation}H_{\text{$10\%$-rule}}(\vec{s}) =  \begin{cases}
    0, & \text{if } \sum_{n=1}^N \delta_{s_n,t} \geq N_t,\\
    p\qty(N_t - \sum_{n=1}^N \delta_{s_n,t}), & \text{else,}
\end{cases}\end{equation}
where $N_t$ is the minimum required number of plies with state $t$.

An intuitive approach to construct a suitable penalty function stems from the following realization: If the constraint is violated, then there exists at least one subset with $N - N_t + 1$ plies, that does not include any plies of state $t$. We can therefore simply define a suitable penalty function by checking all subsets of $N - N_t + 1$ plies and add a penalty, if no plies with state $t$ are present. For example, if we denote the set of all $k$-element subsets of $\{1,\dots,N\}$ with $C(N,k)$, we can define the penalty function as:
\begin{equation}H_{\text{$10\%$-rule}}(\vec{s}) = \sum_{\Lambda \in C(N,N-N_t+1)} \prod_{n \in \Lambda} (1-\delta_{s_n,t}).\end{equation}
Here, $(1-\delta_{s_n,t}) = 1$ exactly if ply $n$ is not in state $t$, $s_n \neq t$, and otherwise the expression is 0. For the terms, we thus add a penalty $\prod_{n \in \Lambda} (1-\delta_{s_n,t}) = 1$, exactly if for all plies $n \in \Lambda$ we have $s_n \neq t$. Since this definition is symmetric under permutations of the plies, and therefore the result only depends on the difference in the number of current and required plies in state $t$, it represents a concrete example of the penalty function in \cref{eq:penalty10perc}. Furthermore, the individual terms translate to operators on $N - N_t + 1$ subsystems in the quantum representation. 

As discussed in \cref{sec:constraints}, these high-order operators can act unfavorably in certain quantum algorithms, especially those that require implementing the Hamiltonian as operations on the qubits. The question now is: Can we do better? As we show in this section, this is unfortunately not the case.

First, we will show, that a penalty function, as defined in \cref{eq:penalty10perc}, will necessarily lead to terms of order $N - N_t + 1$. However, this definition is already somewhat restrictive, as it only allows penalty functions, that are symmetric under the permutation of the plies. From an optimization point of view, it makes sense that states with the same discrepancy to satisfying the constraint produce the same penalty. However, there might be penalty functions, that are not symmetric under permutation, but might be implementable with lower-order operators. In the final part of this section, we will show that any suitable operator for the $10\%$-rule will require operators acting on at least $N - N_t +1$ subsystem, regardless of symmetry considerations.

\subsubsection{Using the defined penalty function in eq. (\ref{eq:penalty10perc})}

In this section, we show, that a decomposition into local operators of the penalty function for the 10\%-rule, as defined in eq. (\ref{eq:penalty10perc}) in section \ref{sec:constraints}:
\begin{equation}H_\mathrm{penalty}^{\text{$10\%$-rule}}(\vec{s}) =  \begin{cases}
    0, & \text{if } \sum_{n=1}^N \delta_{s_n,t} \geq N_t,\\
    p\qty(N_t - \sum_{n=1}^N \delta_{s_n,t}), & \text{else,}
\end{cases}\end{equation}
always includes terms of order $N-N_t+1$.

For this purpose, we extend the definition of $p(x)$ also for valid states, where $x \geq N_t$. If we write $p(x)$ as a polynomial:
\begin{equation}p(x) = \sum_{k=0}^{\deg{p}} a_k x^k,\end{equation}
-- any other form would immediately yield a Taylor expansion up to infinite order -- then each $x \in \{N_t,N_t+1,\dots,N\}$ must be a root of the polynomial:
\begin{equation}p(x) = 0.\end{equation}
Thus, since $p(x)$ has at least $N - N_t + 1$ roots, it must have at least an equivalent degree:
\begin{equation}\deg{p} \geq N - N_t + 1.\end{equation}
For a state $\vec{s} \in \mathcal{S}$, we have:
\begin{equation}x = N_t - \sum_{n=1}^N \delta_{s_n,t},\end{equation}
such that a power $x^k$ yields:
\begin{align}
    x^k &= \qty(N_t - \sum_{n=1}^N \delta_{s_n,t})^k \\
    &= (-1)^k \qty(\sum_{n=1}^N \delta_{s_n,t})^k + (-1)^{k-1} k N_t \qty(\sum_{n=1}^N \delta_{s_n,t})^{k-1} + (-1)^{k-2} \frac{1}{2} (k-1)k N_t^2 \qty(\sum_{n=1}^N \delta_{s_n,t})^{k-2} + \cdots . \notag\\
\end{align}
Furthermore:
\begin{equation}\qty(\sum_{n=1}^N \delta_{s_n,t})^k = \sum_{n_1,\dots,n_k = 1}^{N} \delta_{s_{n_1},t} \delta_{s_{n_2},t} \cdots \delta_{s_{n_k},t}.\end{equation}
Since $p(x)$ contains terms with $k=N-N_t+1$, the penalty function includes terms with $\delta_{s_{n_1},t} \delta_{s_{n_2},t} \cdots \delta_{s_{n_k},t}$, which translate to operators $\ketbra{tt\cdots t}_{n_1 n_2\cdots n_k}$. Therefore, the according Hamiltonian will have terms of order $N-N_t+1$.

\subsubsection{Any penalty function will include high-order terms}

Above, we showed that the specified penalty function in eq. (\ref{eq:penalty10perc}) leads to terms of order $N-N_t+1$. However, the penalty function is inherently symmetric under permutations of the elements in a state $\vec{s}\in\mathcal{S}$. Here, we show that indeed any penalty function, regardless of symmetry considerations, must include terms of order $N-N_t+1$ or larger.

First, we note that non-diagonal local operators produce off-diagonal elements when combined to higher operators. We can therefore focus on diagonal operators. Furthermore, these diagonal operators directly translate to Kronecker-deltas in the classical penalty function, as can be seen from the matrix element of the operator:
\begin{equation}\mel{s}{\qty(\sum_{r = 1}^d c_r \dyad{r})}{s} = c_s = \sum_{r=1}^d c_r \delta_{r,s}\end{equation}
where $r,s \in \{1,\dots,d\}$ are states on a single subsystems. We can therefore focus on expansions of the classical penalty function in terms of Kronecker-deltas.

In order to demonstrate that any penalty function for the $10\%$-rule requires at least $N - N_t + 1$-order terms, we show that any penalty function with lower order will be insufficient. We accomplish this by considering a function that only includes terms of order up to $k = N-N_t$:
\begin{equation}Z(\vec{s}) = \sum_{\Lambda \in \mathcal{P}_{k}(N)} \sum_{\vec{a}\in\{1,\dots,d\}^{|\Lambda|}} \zeta_{\Lambda,\vec{a}} \ \delta_{\vec{s}[\Lambda],\vec{a}},\end{equation}
where $\zeta_{\Lambda,\vec{a}} \in \R$ are the coefficients of the expansion. Here we used the following notation: $\mathcal{P}_k(N)$ is the power set of $\{1,\dots,N\}$, up to cardinality $k$:
\begin{equation}\mathcal{P}_k(N) = \qty{\Lambda \subseteq \{1,\dots,N\} : |\Lambda| \leq k}.\end{equation}
We use subsets $\Lambda \in \mathcal{P}_k(N)$ to select subsequence of elements $s_n$ of the state $\vec{s}$ for $n \in \{1,\dots,N\}$, which we denote with:
\begin{equation}\vec{s}[\Lambda] = (s_{n_1},s_{n_2},\cdots,s_{n_{|\Lambda|}}) \in \{1,\dots,N\}^{|\Lambda|},\end{equation}
where $\Lambda = \{n_1,n_2,\dots,n_{|\Lambda|}\}$ and $n_1 < n_2 < \cdots < n_{|\Lambda|}$. With the Kronecker-delta:
\begin{equation}\delta_{\vec{s}[\Lambda],\vec{a}} = \delta_{s_{n_1},a_1} \delta_{s_{n_2},a_2} \cdots \delta_{s_{n_|\Lambda|},a_{|\Lambda|}} ,\end{equation}
we match the selected states of $\vec{s}$ to a defined vector $\vec{a} \in \{1,\dots,d\}^{|\Lambda|}$. In an operator representation, these Kronecker-deltas are converted into operators $\ketbra{a_1 a_2\cdots a_{|\Lambda|}}_{n_1 n_2 \cdots n_{|\Lambda|}}$.
For our proof, we show that if this function $Z(\vec{s})$ has roots on all valid states, it will necessarily also have roots for at least one constraint-violating state, which makes it unsuitable as a penalty function.

Before we continue, we first note, that:
\begin{equation}\delta_{s_{n},\alpha} =  \delta_{s_{n},\alpha} \sum_{\beta=1}^d \delta_{s_{n'},\beta} = \sum_{\beta=1}^d \delta_{s_{n},\alpha} \delta_{s_{n'},\beta},\end{equation}
from which follows, that terms with lower order can be converted into terms with higher order. Therefore, instead of allowing terms with up to order $k$ in $Z(\vec{s})$, we can focus on $Z(\vec{s})$ containing only terms of exactly order $k$:
\begin{equation} \label{eq:Z_of_s}
    Z(\vec{s}) = \sum_{\Lambda \in C(N,k)} \sum_{\vec{a}\in\{1,\dots,d\}^{|\Lambda|}} \zeta_{\Lambda,\vec{a}} \ \delta_{\vec{s}[\Lambda],\vec{a}},
\end{equation}
where $C(N,k)$ denotes the set of all combinations of $k$ object from $\{1,\dots,N\}$ without replacement:
\begin{equation}C(N,k) = \{\Lambda \in \mathcal{P}(N) : |\Lambda| = k \}.\end{equation}

The valid states $\vec{s}$, with:
\begin{equation}\sum_{n=1}^N \delta_{s_n t} \geq N_t = N-k ,\end{equation}
must not add a penalty:
\begin{equation}Z(\vec{s}) = 0,\end{equation}
which defines a system of linear equations in $\zeta_{\Lambda,\vec{a}}$ in eq. (\ref{eq:Z_of_s}), for which the rows of the according matrix are given by the $\delta_{\vec{s}[\Lambda],\vec{a}}$. We will show, that there exists a state $\vec{r}\in\mathcal{S}$ with less than $N_t$ occurrences of state $t$, and thus should add penalty, but evaluates to $Z(\vec{r})=0$. We do so, by showing, that $\delta_{\vec{r}[\Lambda],\vec{a}}$ can be written as a linear combination of $\delta_{\vec{s}[\Lambda],\vec{a}}$ with valid states $\vec{s}$.

For this purpose, we define vectors $\vec{w}^{(\Gamma)}\in \mathcal{S}$ containing only state $t$ and another state $u \in \{1,\dots,d\}$ with $u \neq t$, where the last $N-k-1$ elements are all $t$:
\begin{equation}\vec{w}^{(\Gamma)} = \vec{x}^{(\Gamma)} \oplus (t,\dots,t). \end{equation}
Here, $\Gamma \in \mathcal{P}(k+1)$ is a subset of $\{1,\dots,k+1\}$ and $\vec{x}^{(\Gamma)} \in \{t,u\}^{k+1}$ is defined by:
\begin{equation}x_n^{(\Gamma)} = \begin{cases}
    t, & \text{if $n\in\Gamma$},\\
    u, & \text{if $n \notin \Gamma$}.
\end{cases}\end{equation}
We will show that $\delta_{\vec{r}[\Lambda],\vec{a}}$ for: 
\begin{equation}\vec{r} = \vec{w}^{(\varnothing)} = (u,\dots,u,t,\dots,t),\end{equation}
which contains $N-k-1 = N_t - 1$ occurrences of $t$, is a linear combination of $\delta_{\vec{w}^{(\Gamma)}[\Lambda],\vec{a}}$ with $\Gamma \neq \varnothing$, such that $\vec{w}^{(\Gamma)}$ constraints at least $N-k = N_t$ occurrences of $t$.

For a given $\Lambda = \{n_1,n_2,\dots,n_k\} \in C(N,k)$ and $\vec{a}\in\{1,\dots,d\}^{|\Lambda|}$, we assume that $n_1,\cdots,n_m \leq k+1$ and $n_{m+1},\cdots,n_k > k+1$. Then for any $\Gamma \in \mathcal{P}(k+1)$:
\begin{equation}\delta_{\vec{w}^{(\Gamma)}[\Lambda],\vec{a}} = \delta_{x^{(\Gamma)}_{n_1},a_1}\delta_{x^{(\Gamma)}_{n_2},a_2}\cdots \delta_{x^{(\Gamma)}_{n_m},a_m} \ \delta_{t,a_{m+1}}\cdots\delta_{t,a_{k}}.\end{equation}
Note that for $\vec{r}$, where $\Gamma=\varnothing$, we get:
\begin{equation} \label{eq:delta_r}
    \delta_{\vec{r}[\Lambda],\vec{a}} =  \delta_{u,a_1} \cdots \delta_{u,a_m} \ \delta_{t,a_{m+1}}\cdots\delta_{t,a_{k}}.
\end{equation}
If we sum the Kronecker-deltas for all $\vec{x}^{(\Gamma)}$ with the same number $\alpha = |\Gamma|$ of $t$, we obtain:
\begin{equation}
    \sum_{\Gamma \in C(k+1,\alpha)} \delta_{\vec{w}^{(\Gamma)}[\Lambda],\vec{a}} = \delta_{t,a_{m+1}}\cdots\delta_{t,a_{k}} \sum_{\Gamma \in C(k+1,\alpha)} \delta_{x^{(\Gamma)}_{n_1},a_1}\delta_{x^{(\Gamma)}_{n_2},a_2}\cdots \delta_{x^{(\Gamma)}_{n_m},a_m}.
\end{equation}
We can evaluate the sum on the right site, by considering that $\vec{x}^{(\Gamma)}$, which has length $k+1$, has to match $\vec{a}$ on the sites defined by $n_1,\dots,n_m$, and that $\gamma = \mathrm{count}\qty(t,(a_1,\dots,a_m))$ occurrences of $t$ are already bound in $\vec{a}$. Therefore, we need to count all combinations, where we place $\alpha - \gamma$ instances of $t$ on the remaining $(k+1) - m$ sites:
\begin{equation}\sum_{\Gamma \in C(k+1,\alpha)} \delta_{\vec{w}^{(\Gamma)}[\Lambda],\vec{a}} = \delta_{t,a_{m+1}}\cdots\delta_{t,a_{k}} \binom{(k+1) - m}{\alpha - \gamma},\end{equation}
where the brackets on the right side denote the binomial coefficient.

Now, we calculate the sum of all these expressions for $\alpha = 1,2,\dots,k+1$ with alternating signs:
\begin{align}
    \sum_{\alpha=1}^{k+1} & (-1)^{\alpha-1}\sum_{\Gamma \in C(k+1,\alpha)} \delta_{\vec{w}^{(\Gamma)}[\Lambda],\vec{a}} \\
    &= \delta_{t,a_{m+1}}\cdots\delta_{t,a_{k}} \sum_{\alpha=1}^{k+1} (-1)^{\alpha-1} \binom{(k+1) - m}{\alpha - \gamma} \notag \\
    &= \delta_{t,a_{m+1}}\cdots\delta_{t,a_{k}} \sum_{\alpha=1-\gamma}^{k+1-\gamma} (-1)^{\alpha + \gamma-1} \binom{(k+1) - m}{\alpha} \notag \\
    &= \delta_{t,a_{m+1}}\cdots\delta_{t,a_{k}} (-1)^{\gamma-1} \sum_{\alpha=\max(0,1-\gamma)}^{k+1-m} (-1)^{\alpha} \binom{(k+1) - m}{\alpha}, \notag
\end{align}
where in the second line, we shifted the index $\alpha$ and in the third line, we trimmed to index $\alpha$ to reflect, that the binomial coefficient vanishes for $\alpha < 0$ or $\alpha > (k+1)-m$. We distinguish two cases: If the number of $t$ in $(a_1,\dots,a_m)$ is non-zero, then the expression vanishes:
\begin{align}
    \sum_{\alpha=1}^{k+1} & (-1)^{\alpha-1}\sum_{\Gamma \in C(k+1,\alpha)} \delta_{\vec{w}^{(\Gamma)}[\Lambda],\vec{a}} \\
    &= \delta_{t,a_{m+1}}\cdots\delta_{t,a_{k}} (-1)^{\gamma-1} \sum_{\alpha=0}^{k+1-m} (-1)^{\alpha} \binom{(k+1) - m}{\alpha}\notag\\
    &=0,\notag
\end{align}
where we used the identity $\sum_{j=1}^{n} (-1)^j \binom{n}{j} = 0$. If $(a_1,\dots,a_m)$ does not contain any $t$, then we need to make sure to include the term $\alpha = 0$, in order to use this identity:
\begin{align}
    \sum_{\alpha=1}^{k+1} & (-1)^{\alpha-1}\sum_{\Gamma \in C(k+1,\alpha)} \delta_{\vec{w}^{(\Gamma)}[\Lambda],\vec{a}} \\
    &= \delta_{t,a_{m+1}}\cdots\delta_{t,a_{k}} (-1)^{\gamma-1} \sum_{\alpha=1}^{k+1-m} (-1)^{\alpha} \binom{(k+1) - m}{\alpha},\notag\\
    &= \delta_{t,a_{m+1}}\cdots\delta_{t,a_{k}} (-1)^{\gamma-1} \qty(\sum_{\alpha=0}^{k+1-m} (-1)^{\alpha} \binom{(k+1) - m}{\alpha} - (-1)^0 \binom{(k+1) - m}{0})\notag\\
    &= \delta_{t,a_{m+1}}\cdots\delta_{t,a_{k}},\notag
\end{align}
where in the last line, we included that the number of $t$ in $(a_1,\dots,a_m)$ vanishes in this case, $\gamma=0$ and thus $(-1)^{\gamma-1}=-1$.
We can summarize the results for the two cases as:
\begin{equation}\sum_{\alpha=1}^{k+1} (-1)^{\alpha-1}\sum_{\Gamma \in C(k+1,\alpha)} \delta_{\vec{w}^{(\Gamma)}[\Lambda],\vec{a}} = \delta_{u,a_1} \cdots \delta_{u,a_m} \ \delta_{t,a_{m+1}}\cdots\delta_{t,a_{k}}, \end{equation}
which is exactly $\delta_{\vec{r}[\Lambda],\vec{a}}$ in eq. (\ref{eq:delta_r}):
\begin{equation}\delta_{\vec{r}[\Lambda],\vec{a}} = \sum_{\alpha=1}^{k+1} (-1)^{\alpha-1}\sum_{\Gamma \in C(k+1,\alpha)} \delta_{\vec{w}^{(\Gamma)}[\Lambda],\vec{a}}. \end{equation}
Therefore $\delta_{\vec{r}[\Lambda],\vec{a}}$ is linear combination of $\delta_{\vec{w}[\Lambda]^{(\Gamma)},\vec{a}}$. As a consequence, if we choose $\zeta_{\Lambda,\vec{a}}$ in eq. (\ref{eq:Z_of_s}), such that $Z(\vec{w}^{(\Gamma)}) = 0$ for all valid states with $\Gamma \neq \varnothing$, then $Z(\vec{r}) = 0$ for the constraint violating state $\vec{r}$. Thus, a suitable penalty function $Z(\vec{s})$ can not be constructed with only terms of order less then $N-N_t + 1$. As a consequence, any penalty Hamiltonian will contain terms of order $N - N_t + 1$ or more.

\subsection{Symmetries of stacking sequences with respect to lamination parameters and constraints} \label{sec:app_sym}

In \cref{sec:demo}, we selected target lamination parameters for $N=6$ plies that correspond to stacking sequences which cannot be transformed into each other through symmetry transformations of the lamination parameters and the disorientation constraint. In this section, we explore these symmetry transformations and their impact on SSR.

Given their definition, the set of lamination parameters is symmetric under specific rotations and reflections of the space. As long as an SSR algorithm inherently treats all ply-angles equally, it can be expected to perform similarly for target lamination parameters that can be transformed into each other through these symmetry transformations. For numerical trials of the algorithm, it therefore makes sense to choose target parameters that are distinctly invariant to ensure a broad representation of potential behaviors. We begin by examining the symmetries of the lamination parameter space with arbitrary continuous ply-angles and then discuss the scenario with discrete parameters.

The 4-dimensional rotation:
\begin{equation} R(\chi) = \begin{pmatrix}
    \cos(2 \chi) & - \sin(2 \chi) & 0 & 0 \\
    \sin(2 \chi) & \cos(2 \chi) & 0 & 0 \\
    0 & 0 & \cos(4 \chi) & - \sin(4 \chi)\\
    0 & 0 & \sin(4 \chi) & \cos(4 \chi)
\end{pmatrix}  \end{equation}
applied to the set of lamination parameters of a ply-angle sequence $\vec{\phi} = (\phi_1,\dots,\phi_N)$ for a specific $X=A,B,D$:
\begin{equation} \begin{pmatrix} v^X_1 \\ v^X_2 \\ v^X_3 \\ v^X_4 \end{pmatrix} = \sum_{n=1}^N \alpha^X \begin{pmatrix}
    \cos(2 \phi_n) \\ \sin(2 \phi_n) \\ \cos(4 \phi_n) \\ \sin(4 \phi_n)
\end{pmatrix}  \end{equation}
produces a constant shift of $\chi$ for all ply angles $\phi_n$:
\begin{equation} R(\chi) \begin{pmatrix} v^X_1 \\ v^X_2 \\ v^X_3 \\ v^X_4 \end{pmatrix} = \sum_{n=1}^N \alpha^X \begin{pmatrix}
    \cos(2 (\phi_n+\chi)) \\ \sin(2 (\phi_n+\chi)) \\ \cos(4 (\phi_n+\chi)) \\ \sin(4 (\phi_n+\chi))
\end{pmatrix}.  \end{equation}
As the shifted angle sequence $(\phi_1+\chi,\dots,\phi_N+\chi)$ remains a valid ply angle sequence, the set of lamination parameters is symmetric under performing rotation $R(\chi)$ on all $X=A,B,D$ simultaneously with the same angle $\chi$.

Additionally, the set of lamination parameters possesses mirror symmetry by applying the reflection:
\begin{equation} T = \begin{pmatrix}
    1 & 0 & 0 & 0 \\
    0 & -1 & 0 & 0\\
    0 & 0 & 1 & 0 \\
    0 & 0 & 0 & -1
\end{pmatrix} \end{equation}
to all $X=A,B,D$ simultaneously, which coincides with changing the sign of the ply angles:
\begin{equation} T \begin{pmatrix} v^X_1 \\ v^X_2 \\ v^X_3 \\ v^X_4 \end{pmatrix} = \sum_{n=1}^N \alpha^X \begin{pmatrix}
    \cos(-2 \phi_n) \\ \sin(-2 \phi_n) \\ \cos(-4 \phi_n) \\ \sin(-4 \phi_n)
\end{pmatrix}  \end{equation}
Given that for $\chi \in (90^\circ, 90^\circ]$ the rotations $R(\chi)$ are distinct, the group generated by the rotations $R(\chi)$ and the reflection $T$ is isomorphic to the orthogonal group $O(2)$. Note that here the rotations have a periodicity of $180^\circ$.

Next, we examine the case of $d$ evenly-spaced discrete angles $\Theta = \{0^\circ, \Delta \theta, 2 \Delta \theta, \dots, 180^\circ - \Delta \theta\}$ where $\Delta \theta = \frac{180^\circ}{d}$. This includes the commonly used four angles $\{0^\circ, +45^\circ, 90^\circ, -45^\circ\}$, with $-45^\circ$ being equivalent to $135^\circ$. In this setting, the shift $\chi$ induced by a rotation $R(\chi)$ must also produce a valid ply-angle sequence, thus $\chi$ is restricted to the set $\Theta$. The reflection $T$ continues to be a symmetry transformation of the lamination parameters. With these restrictions, the group generated by the permissible rotations and the reflection forms a subgroup of the previous group, isomorphic to the dihedral group $D_d$, which is the symmetry group of a polygon with $d$ vertices.

The symmetry transformations may be further constrained by manufacturing requirements. For the disorientation constraint, shifting all ply-angles does not alter the ply-angle difference between two plies; thus, a valid ply-angle sequence remains valid under these transformations, while a constraint-violating state remains invalid. However, this is not generally the case for the balanced laminates constraint. For instance, requiring an equal number of $+45^\circ$ and $-45^\circ$ plies, a rotation of $45^\circ$ swaps these angles with $0^\circ$ and $90^\circ$, which may result in a ply-angle sequence alternating between valid and invalid states. Consequently, an SSR algorithm may perform differently under transformations that are not symmetries of the constraint.

In our simulations involving the four angles $\theta_1 = 0^\circ, \theta_2 = +45^\circ, \theta_3 = 90^\circ, \theta_4 = -45^\circ$ and only a disorientation constraint, the transformations of stacking sequences $\vec{s} = (s_1,\dots,s_N) \in \mathcal{S}$ corresponding to allowed rotations can be expressed as adding an integer modulo 4 to all ply-states $s_n$. Possible reflections include swapping states 1 and 3, or states 2 and 4. For the selected target lamination parameters for $N=6$, these stacking sequences cannot be transformed into each other using any combination of these transformations.

At the beginning of this section, we noted that an SSR algorithm is expected to behave similarly for target parameters that are equivalent under these symmetry transformations, provided it does not discriminate between different ply angles. This is certainly the case for the DMRG algorithm, which directly operates on the ply-states. For the QAOA algorithm and the hardware-efficient approach, consideration must be given to the qubit-encoding, the parameterized quantum circuit, and its optimization. In our state-vector simulations, qubit states $\ket{0}$ and $\ket{1}$ are considered interchangeable, but on actual quantum hardware, these states often do not exhibit completely symmetric behavior. For instance, one of these states may be a higher energy state prone to erroneously falling into the lower energy state ($T_1$ decay), potentially leading to performance differences between encodings that involve different durations of the qubits being in the higher energy state.

Moreover, even within state-vector simulations, performance differences may occur for equivalent target parameters. The QAOA algorithm applies operations to all qubits simultaneously and implements eigenvalue-dependent phases, which should make it less susceptible to encoding-specific features. However, the local optimizations in the hardware-efficient approach may exhibit a dependency on the encoding, potentially resulting in distinct behaviors for target parameters that can be transformed into each other. For this reason, we have attempted to incorporate the rotational symmetries into our encoding at least partially. In this encoding, for the qubits representing one ply, flipping one qubit results in a rotation of $+45^\circ$ or $-45^\circ$, while flipping both qubits results in a rotation of $90^\circ$. This remains consistent even under the aforementioned symmetry transformations.

\subsection{Diagram of all measured samples for DMRG with \texorpdfstring{$N=200$}{N=200} plies} \label{sec:app_allsamples}

\begin{sidewaysfigure}
    \centering
    \includegraphics[scale=1.0]{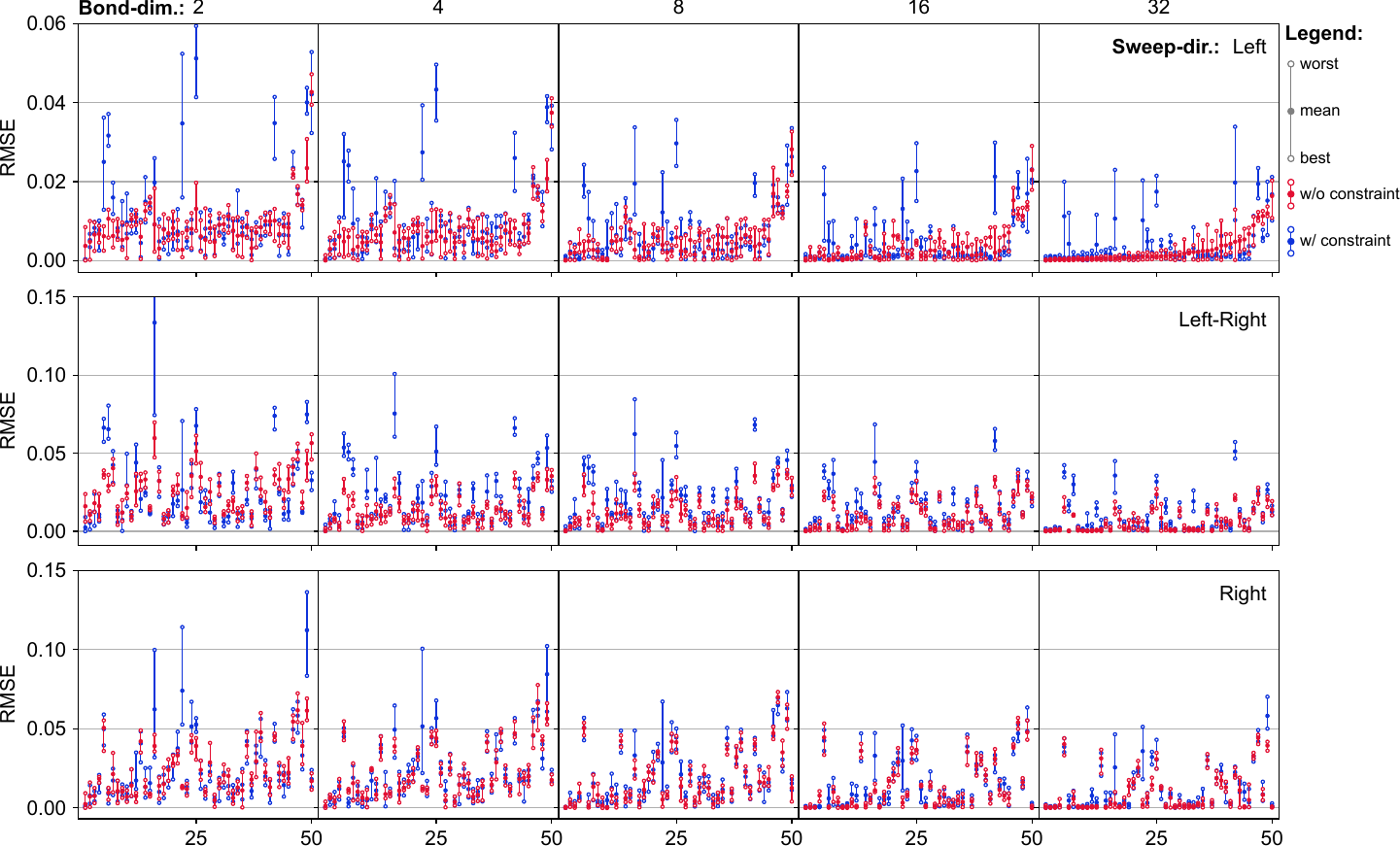}
    \caption{Summary of all DMRG samples organized in a tabular format. Rows correspond to different sweep directions (inward, alternating, outward) and columns represent the bond dimensions ranging from 2 to 32. For each sample, the best, worst, and average results are shown. The samples are sorted by the average values for bond dimension 32 and sweep direction left, without the constraint (top right).}\label{fig:dmrg_samples}
\end{sidewaysfigure}

Figure \ref{fig:dmrg_samples} offers a tabular depiction of the outcomes of all measured samples. The rows and columns respectively represent different sweep directions and diverse bond dimensions.  Here, the rows signify different sweep directions, and the columns represent the diverse bond dimensions, ranging from 2 to 32. The samples are sorted by the average values for bond dimension 32 and sweep direction left, without the constraint.

